\documentclass[a4paper,10pt]{article}
\usepackage{lineno}
\usepackage[utf8]{inputenc}
\usepackage{amsmath}
\usepackage{graphicx}
\usepackage{lscape}
\usepackage{fullpage}
\usepackage[authoryear,sectionbib,sort]{natbib}
\usepackage{multirow}
\usepackage{latexsym, amsfonts, amssymb, amsthm, amscd}
\usepackage{rotating}
\usepackage{subfig}
\usepackage{comment}

\usepackage{longtable}
\usepackage{titlesec}
 \titleformat{\section}[block]{\Large\bfseries\filcenter}{\thesection}{1em}{}
 \titleformat{\subsection}[block]{\Large\itshape\filcenter}{\thesubsection}{1em}{}
 \titleformat{\subsubsection}[block]{\large\itshape}{\thesubsubsection}{1em}{}
 \titleformat{\paragraph}[runin]{\itshape}{\theparagraph}{1em}{}[. ]
\usepackage{authblk}
\usepackage[table]{xcolor}  
 \usepackage{marvosym}
\usepackage{natbib}
\usepackage{fancybox}
\newcommand{\eg}{\emph{e.g.}}
\newcommand{\ie}{\emph{i.e. }}

\title{\bf Stochastic dynamics of three competing clones: Conditions and times for invasion, coexistence and fixation}
\author[1]{}
\author[1]{Sylvain Billiard \thanks{Corresponding Author, sylvain.billiard@univ-lille.fr}}
\author[2,3]{Charline Smadi \thanks{charline.smadi@irstea.fr}}
\affil[1]{ Unité \'Evo-\'Eco-Paléo, UMR CNRS 8198, Université des Sciences et Technologies Lille, 59655 Villeneuve d'Ascq Cedex, France}
\affil[2]{ Universit\'e Clermont Auvergne, Irstea, UR LISC, Centre de Clermont-Ferrand, F-63178 Aubi\'ere, France}
\affil[3]{Institut des Syst\`emes Complexes Paris Ile-de-France, 113 rue Nationale, 75013 Paris, France.}
\bigskip

\bigskip

\begin{document}

\maketitle


\par
The authors wish to be identified to the reviewers.

\newpage{}
%\normalem
\section*{Abstract}
In large clonal populations, several clones generally compete which results in complex evolutionary and ecological dynamics: 
experiments show successive selective sweeps of favorable mutations as well as long-term coexistence of multiple clonal strains. 
The mechanisms underlying either coexistence or fixation of several competing strains have rarely been studied altogether. Conditions 
for coexistence have mostly been studied by population and community ecology, while rates of invasion and fixation have mostly been studied 
by population genetics. In order to provide a global understanding of the complexity of the dynamics observed in large clonal populations, 
we develop a stochastic model where three clones compete. Competitive interactions can be intransitive and we suppose that strains enter 
the population via mutations or rare immigrations. We first describe all possible final states of the population, including stable 
coexistence of two or three strains, or the fixation of a single strain. Second, we give estimate of the invasion and fixation times of 
a favorable mutant (or immigrant) entering the population in a single copy. We show that invasion and fixation can be slower 
or faster when considering complex competitive interactions. Third, we explore the parameter space assuming prior distributions of 
reproduction, death and competitive rates and we estimate the likelihood of the possible dynamics. We show that when mutations can affect competitive interactions, 
even slightly, stable coexistence is likely. We discuss our results in the context of the evolutionary dynamics of large clonal 
populations.
\newpage

\section*{Introduction}\label{sec:intro}

When different genotypes, strains, or species compete within a population, two categories of outcomes can be expected at first sight: either several types of individuals stably coexist or a single type goes to fixation. However, evolution experiments of large clonal populations show more complex dynamics, even in well-mixed environments, with a succession of different phases: several favorable mutations can simultaneously compete, some of them can go to fixation while others go extinct, and several clones can coexist for a long time, sometimes with apparent cyclical dynamics \citep{hellingetal87, raineyandtravisano1998,langetal11, maddamsettietal15, behringeretal2018, levyetal2015}. Despite a large theoretical ecological and evolutionary literature dealing with the dynamics of clones communities, the complexity of evolutionary dynamics observed in evolution experiments is still largely unexplained. Our goal is to show that bridging the gap between population ecology and population genetics models can give new insights on the mechanisms underlying the evolutionary dynamics of large clonal populations.

Through the lens of population genetics, most investigations focused on how the co-occurrence of several favorable mutations in a single population would 
affect adaptation, and especially adaptation rates, \ie on the speed at which a population can adapt to a given environment due to the 
fixation of beneficial mutations \citep{fisher1930, muller32}. Probability and time of invasion and fixation of rare mutants 
have been
thoroughly investigated in population and evolutionary genetics for decades \cite[\eg][]{crowkimura1965}. When adaptation is due to successive selective sweeps,  with at most two competing strains at a given time, 
adaptation rate is proportional to population size, the mutation rate and the effect of mutation on fitness. This is however not true in large clonal populations: evolution experiments with  micro-organisms show that multiple strains compete in a population at a given time, even though they were started with a single strain. In large clonal populations, 
mutations arise at a rate higher than they invade and go to fixation, resulting in a large number of competing clones. 
It has dramatic consequences on adaptation rate, which is generally slower: it is proportional to a power of the logarithm of the population size, a phenomenon generally called ``clonal interference'' or ``concurrent mutations regime'' \citep{gerrishlenski98, neher13}. Clonal interference has been observed in  bacteria, viruses, yeasts or cancer tumors \citep[\eg][]{mirallesetal99, devisserrozen06,  hegrenessetal2006, greavesmaley12, langetal13, levyetal2015}. The relationship between adaptation rates and mutation rates and selection also depends on whether favorable mutations appear on different lineages or not \citep{parketal2010, desaifisher07, goodetal12}.

Even though the theory 
about invasion and fixation rates when multiple clones compete is advanced in population genetics literature, it is based on important simplifying assumptions: population size is a fixed parameter, and mutations are supposed to have transitive effects, \ie mutations only affect the reproduction rates of clones, and not the competitive interactions between individuals within or between clonal strains. Such assumptions hinder the possibility of stable coexistence of several strains in a single population in a well-mixed homogeneous environment where long-term coexistence and non-linear dynamics are common \citep{langetal11, maddamsettietal15, goodetal2017, behringeretal2018}. Unpredicted evolutionary dynamics has been explained by varying mechanisms. \citet{langetal11} observed different replicates of yeasts evolution experiments where a lineage showed two successive frequency peaks, that they explained by the occurrence of a third cryptic mutation affecting a preexisting lineage. Several other experiments showed  coexistence of different clonal strains in the long-term, what was interpreted as an evidence of frequency-dependent selection \citep{raineyandtravisano1998, maharjanetal06,langetal11, maddamsettietal15, goodetal2017, behringeretal2018}. \citet{goodetal2017} in particular showed that the observed long-term coexistence 
of different clonal strains cannot be due to clonal interference only. \citet{rosenzweigetal1994} and \citet{kinnersleyetal14} showed the long-term coexistence of three lineages derived by mutation from a single initial \emph{Escherichia coli} clone in a chemostat what they explained with cooperative rather than competitive interactions. Even if a large literature deals with balancing selection in population genetics literature \citep[see][for a review]{llaurensetal2017}, to what extent frequency-dependent selection or non-transitive interactions can affect evolutionary dynamics and adaptation of large clonal populations has received few attention.

Many models from the ecological literature looked for the mechanisms promoting or not coexistence of several species or strains \citep[reviewed in][]{chesson2000, chesson2018}. It is now well established that stable coexistence depends on how within- and inter-species competitions relate \citep{chesson2000, barabasetal2016}, with involved mechanisms such as competition for different resources \citep{tilman1982, goodetal2018}, spatialized interactions \citep[\eg][]{pacalaandtilman1994, vetsigian2017}, environmental filtering \citep{ackerlyandcornwell2007}, or intransitive competitive interactions \citep{gallienietal2017}. Intransitive competition occurs between three species A, B and C when A is a better competitor than B and B than C but A is not better than C. Intransitive competitive interactions are of particular interest in the context of large clonal populations since evolution experiments with micro-organisms show complex dynamics even in uniform environments with a few shared resources (often a single one).  Different mechanisms can underlie intransitivity for competitive interactions: trade-offs, life-history traits change between developmental stages,  variability in efficiency in the use of different resources, or space \citep[reviewed in ][]{gallieni2017}. Experimental studies showed that intransitive competitive interactions commonly occur in plants, bacteria, fungi, protists, corals or lizards, in some cases both within and between species  \cite[\eg][]{taylorandaarssen1990, nahumetal2011, abelsonandloya1999, sinervoandlively1996, friedmanetal2017, gallieni2017, soliveresetal2018}.

Theoretical works showed that intransitive  competitive interactions can promote coexistence, even in the simplest models  \cite[\eg \ the Lotka-Volterra competitive model, ][]{mayandleonard1975, zeemanzeeman03, gallienietal2017}. However, all theoretical works studying the effect of competitive interactions on coexistence, including intransitivity, used two types of criteria:  either the invasibility of a species when all others are at equilibrium \cite[\eg][]{doebeli2002, gallienietal2017, goodetal2018} or the stability of the equilibrium of a community of species \cite[\eg][]{barabasetal2016}. These models have three important drawbacks. First, they implicitly assume a large number of individuals of the different species which, on the one hand, hinders considering fixation as a possible final state and, on the other hand, does not give insights on the probability of invasion of a mutant or immigrant.  Second, as shown by the evolution experiments, there is no particular 
reason why the community should be at equilibrium when a mutant or an immigrant enters into a population. Investigating the conditions for the invasion of a rare species or clonal 
strain assuming that the community is at equilibrium might thus not capture the whole complexity of the dynamics. For instance, the time at which a mutant or immigrant enters 
a population can dramatically affect the final states of the population. Third, they generally do not consider the particular case of a community derived from a single strain by mutations \cite[see however][]{goodetal2018}. Hence, in order to fully understand the conditions for stable coexistence, including the conditions for 
the establishment of such a coexistence with rare mutants or immigrants, it is necessary to use stochastic models.  In addition, ecological models do not generally consider invasion and fixation times of favorable mutants, they thus cannot inform us about adaptation rates.

In summary, population genetics models on the one hand, and population ecology models on the other hand are not general enough to embrace all dynamics observed in experiments, and bridging the gap between both is needed to better disentangle the mechanisms underlying the complexity of the evolution of clonal populations. Such a goal is particularly relevant here because it seems unclear which one of species sorting or mutant fixation by natural selection should better explain the evolutionary dynamics of large clones communities. In this paper, we develop and analyse a stochastic model where multiple clones can compete, assuming density-dependent competition affecting death rates, and where competitive interactions can be transitive or not. Our main goal is to evaluate to what extent intransitive competitive interactions can affect and explain observed evolutionary dynamics. We give explicit results for a short time scale, and we discuss the implications of our results for large time scales evolution experiments. In a previous paper, we demonstrated that the stochastic dynamics of multiple competing clones can be approximated by a succession of branching processes without 
interactions and Lotka-Volterra deterministic systems \citep{billiardsmadi17}. In the present paper, we build upon \citet{billiardsmadi17}. We focus on the simple case with 
only three competing clones. Our objectives are 1) To give general conditions under which different final states are obtained depending on the ecological parameters and on when 
clones enter the population; 2) To give approximations of the times of invasion and fixation of favorable mutations when three clones compete as a function of population size and 
mutation effect on fitness; 3) To explore the parameter space and determine the likelihood of the different possible dynamics and final states by assuming prior distributions 
on ecological parameters. We show in particular that our model captures a large variety of dynamics and patterns observed in evolution experiments. We also show that fixation 
of favorable mutations can go slower or faster depending on the competitive interactions and on the time when mutations enter the population. Finally we show that when 
mutations can affect competitive interactions, even with very small effect, stable coexistence is likely. We conclude by arguing that our present work is an illustration that 
theoretical frameworks from population genetics and population ecology can be gathered in order to have a broader understanding of evolutionary and ecological dynamics.

\section*{A stochastic model, its approximations and properties}\label{sec:model}

\subsection*{The model.} We consider that the population is composed of clonal individuals, with different possible strains denoted $i$.  For the sake of simplicity, we will only use the term 
\emph{clone} $i$ to refer to type $i$ individuals ($i$ could however refer to different phenotypes, alleles, strains, clonal species, lineages, mutants, etc.).  
We denote $N_i(t)$ the number of clone $i$ individuals in the population at time $t$ ($N_i(t)$ is a random variable). We investigate the population dynamics of different 
competing clones as a birth-death process with competition in continuous time (see Tab. \ref{tab:summary} for a summary of parameters and variables used in the model). Each clone $i$ individual is characterized by its ecological parameters: $\beta_i$ and $\delta_i$ are the individual birth and natural death rates, respectively. The effect of competition of a single clone $j$ individual on a single clone $i$ individual, denoted $C_{ij}$, is assumed to affect mortality only, adding a component $\sum_j C_{ij} N_j(t)/K$ to individual $i$ death rate, with $K$ a scaling parameter. The total individual death rate of clone $i$ thus 
depends on both an intrinsic component ($\delta_i$) and a competition component: $d_i(N(t))=\delta_i+\sum_j C_{ij} N_j(t)/K$. Since we want competitive interactions to increase mortality, we assume that  $C_{ij} \geq 0$ and $C_{ii}>0$. 
Note that depending on the values of the $C_{ij}$, competitive interactions can be transitive or not. 
At any time $t$, given the composition of the population $N_i(t)$ for all $i$, different events can occur at next time step $t+\Delta t$: either the death or a birth of a clone $i$ individual. The probability of each event is given by the ratio between its rate and the total event rate (note that $\Delta t$ is a random variable following an exponential distribution with the total event rates as a parameter, see App. A2).  Hence, the birth or death of a clone $i$ individual occurs with probabilities given by

\begin{align*}
  P(\text{birth of an individual $i$}|N_0(t), ...) & = \frac{\beta_i N_i(t)}{ \sum_{j} N_j(t) \left( \beta_j+\delta_j+\sum_{k}C_{kj}N_k(t)/K \right)},\\
  P(\text{death of an individual $i$}|N_0(t), ...) & = \frac{\delta_i N_i(t)+ \sum_{k}C_{ki}N_k(t)N_i(t)/K}{\sum_{j} N_j(t) \left( \beta_j+\delta_j+\sum_{k}C_{kj}N_k(t)/K \right)}.
\end{align*}\\

 Because the rate of each event depends non-linearly on the number of individuals, the stochastic dynamics cannot be entirely described. However, approximations can be used to highlight the different possible dynamics and final states. We showed earlier that the dynamics of multiple competing clones  can be decomposed into successive phases well approximated either by a deterministic 
Lotka-Volterra model or by a stochastic branching process \citep{billiardsmadi17}. In the present paper we 
build upon the mathematical proofs derived by \cite{billiardsmadi17} to show the implications of multiple competing clones for adaptation and clonal species invasion, loss or 
coexistence. In order to provide analytical predictions supported by stochastic simulations, we will focus on the cases with only three competing clones, for the sake 
of simplicity. We will i) show that multiple dynamics and final states are possible depending on mutational effects on fitness 
and the time separating mutations;  ii) give precise approximations of invasion and fixation times, and invasion probability; iii) explore the likelihood of the different dynamics and 
final states depending on prior distributions on parameters. As shown by \cite{billiardsmadi17}, the dynamics of four or more competing clones can similarly be decomposed 
into a succession of phases. However, the dynamics and final states are difficult to predict since a deterministic Lotka-Volterra with more than three clones can show chaotic dynamics \citep{vanoetal06,wangxiao10}.

\subsection*{Three different regimes.} The number of different competing clones at a given time in the population depends on the relative values of ecological and evolutionary parameters such as the mutation rate and selection, and the scaling parameter $K$ which gives the intensity of stochasticity (or, analogously, genetic drift). Excluding the trivial case with a single clone, three different regimes can be considered: only two, a few, or many competing 
clones. Each regime is characterized by the balance between the time separating two successful favorable mutations and the duration of a successful invasion. To be more 
precise, let us consider favorable mutations with invasion fitness $S$, arising by mutation at individual rate $\mu$ in a population with size of order $K$ and where individuals 
reproduce at rate $\beta$ (a precise definition of $S$ will be given later). Favorable mutations enter the population at rate $K \ \mu$ and each 
mutation has a probability $S/\beta$ to reach a critical size and not be lost by chance. This yields that the time separating the invasion of two favorable mutations is of 
order $1/ (K \ \mu \ S/\beta)$ \citep{fourniermeleard04, desaifisher07, neher13}. The time taken for a favorable mutation to invade the population is approximately given by 
$\ln(K S/\beta) /S$ \citep{desaifisher07, neher13}. 

\begin{figure}[p!]
\begin{center}
\includegraphics[trim = 10mm 70mm 1mm 0mm, clip,scale=0.6]{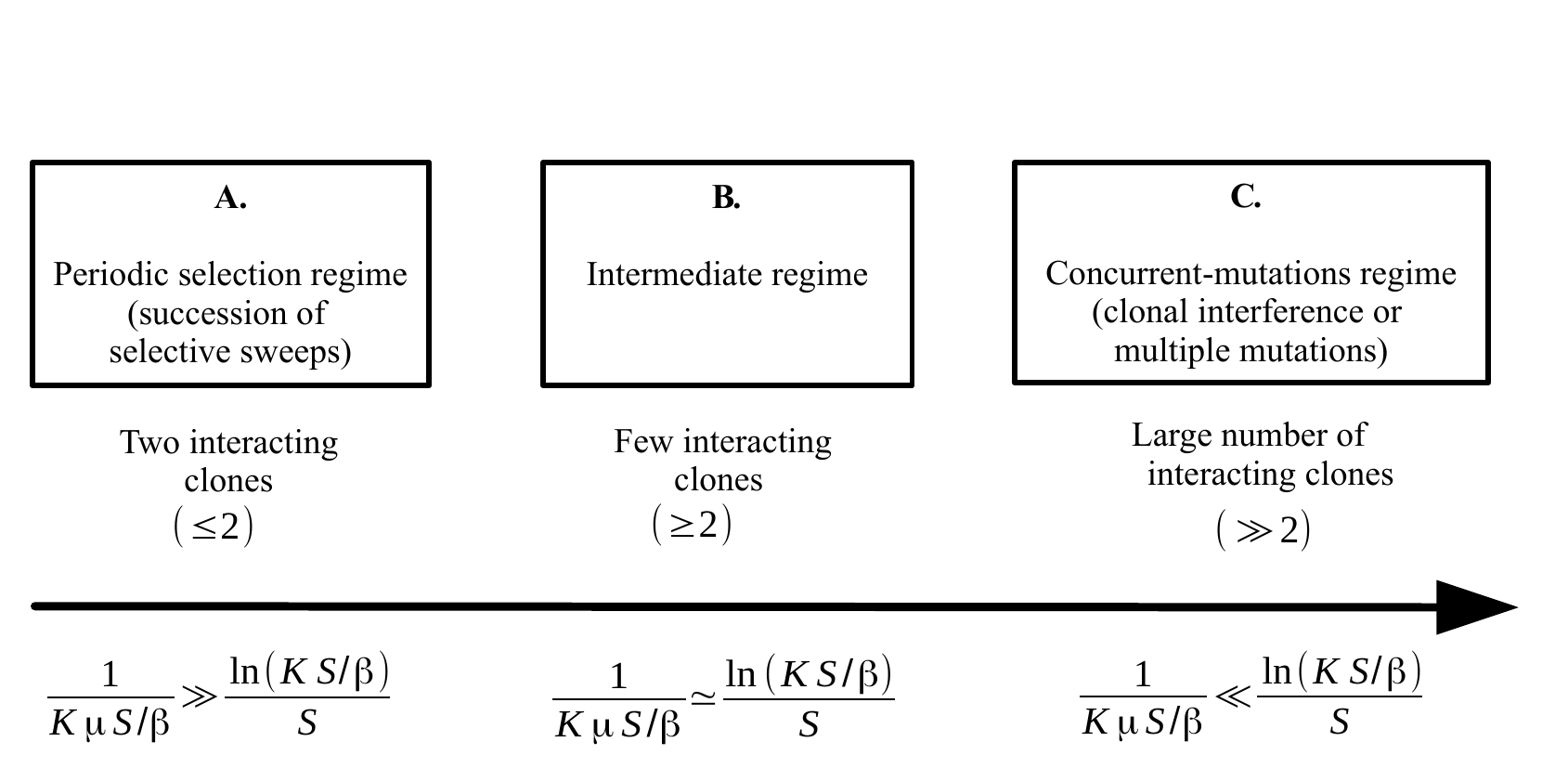}
\end{center}
\caption{The different regimes of mutation-selection in finite clonal populations. The number of clones competing in a population at a given time depends 
on the rate of favorable mutations $\mu$, the effect of mutations on fitness $S$, population size $K$ and individual reproduction rate $\beta$ 
(see text for details). {\bf A.} Periodic selection regime: at most two different clones compete. {\bf B.} Intermediate regime: a few competing clones. 
{\bf C.} A large number of competing clones.}
\label{fig:regime}
\end{figure}

Comparing $1/ (K \ \mu \ S/\beta)$ and  $\ln( K S/\beta) /S$ gives quantitative conditions for the three possible regimes presented above (Fig. \ref{fig:regime}). 
If $1/ (K \ \mu \ S/\beta) \gg \ln(K S/\beta) /S$, the time between two favorable mutations is much higher than the fixation time. In this regime, under the assumption that 
mutation effects on fitness are transitive, favorable mutations get fixed successively in the population, resulting in a succession of selective sweeps 
(called \emph{periodic selection regime} \citep{barricklenski13} or \emph{trait substitution sequence} \citep{champagnat06}). More generally, the stochastic dynamics of 
several competing clones, including non-transitive interactions between clones, have been studied in \cite{champagnatmeleard11}. However, \citet{champagnatmeleard11} assumed 
a timescale separation, \ie new clonal strains enter the population under the hypothesis that the resident population is at a steady state. Such an assumption precludes investigating the dynamics of several clones at a given time, and does not give information about the time of invasion and fixation of clones.  
If $1/ (K \ \mu \ S/\beta) \ll \ln(K S/\beta) /S$, favorable mutations enter the population much faster than they get fixed, which results in the coexistence of many competing 
clones  (called \emph{clonal interference, concurrent mutations}, or \emph{multiple-mutations} regime depending on the assumptions about the amplitude of mutational effects 
and their underlying mechanisms \citep{gerrishlenski98, desaifisher07,  neher13}). The regime where many clones compete has been studied more recently under the assumption 
that mutational effects on fitness are transitive \cite[see][for a review]{neher13}. Under this regime, the rate of adaptation increases sub-linearly with 
population size. Finally, when $1/ (K \ \mu \ S/\beta) \simeq \ln(K S/\beta) /S$, an intermediate regime can be expected, where only a few clones compete at a given time in the 
population, two or more. 

In the present paper, we assume being in this intermediate regime where a few clones compete. Only a few clones compete when the beneficial mutation rate is not too high ($1/K\mu  \simeq \ln K$, Fig. \ref{fig:regime}) and the effect of mutation on fitness is not too low relatively to population size ($S \gg 1/K$).  We will investigate the dynamics and final states of a small 
clonal community, and especially the impact of intransitive competitive interactions (depending on the sign of the $C_{ij}$'s) and the time at which clones enter the population.   Because we assume being in an intermediate regime with the mutation rate such as $1/K\mu  \simeq \ln K$ (Fig. \ref{fig:regime}), time is measured in units of $\ln K$. Hence, we assume that clones enter the population at time $\alpha \ln K$.  We will only consider dynamics on a short time scale relative to the mutation rate, \ie we assume no recurrent mutations affecting fitness entering the population. Consequently we will not investigate the rate of adaptation as clonal interference models did  \citep{gerrishlenski98, desaifisher07,  neher13}. We rather focus on determining the conditions for invasion, fixation and coexistence of different clones, and on the dynamics durations. Understanding what happens on short time scales is indeed necessary to explain complex dynamics observed in evolution experiments on large time scales. 

%4. Comment définir alpha en dehors de alpha Ln K ? On se place dans le régime intermédiaire, c'est à quand 1/KU .. \simeq Ln K ... (Fig. 1) : le temps moyen d'apparition de nouvelles mutaitions est 1/KU, et si on veut être dans le bon régime il faut qu'on fasse l'hypoth_se que ce temps est du bon ordre, c'est à dire que 1/ Ku ~ Ln K, et du coup on pose alpha Ln K comme ce temps.

\subsection*{Approximations of the stochastic dynamics as a succession of phases}
\begin{figure}
\begin{center}
%\subfloat{
%\includegraphics[trim = 0mm 5mm 0mm 20mm, clip, scale=0.69]{dyn2clones.pdf} \label{subfig:Fix2}}\\
\includegraphics[trim = 0mm 5mm 0mm 20mm, clip,scale=1]{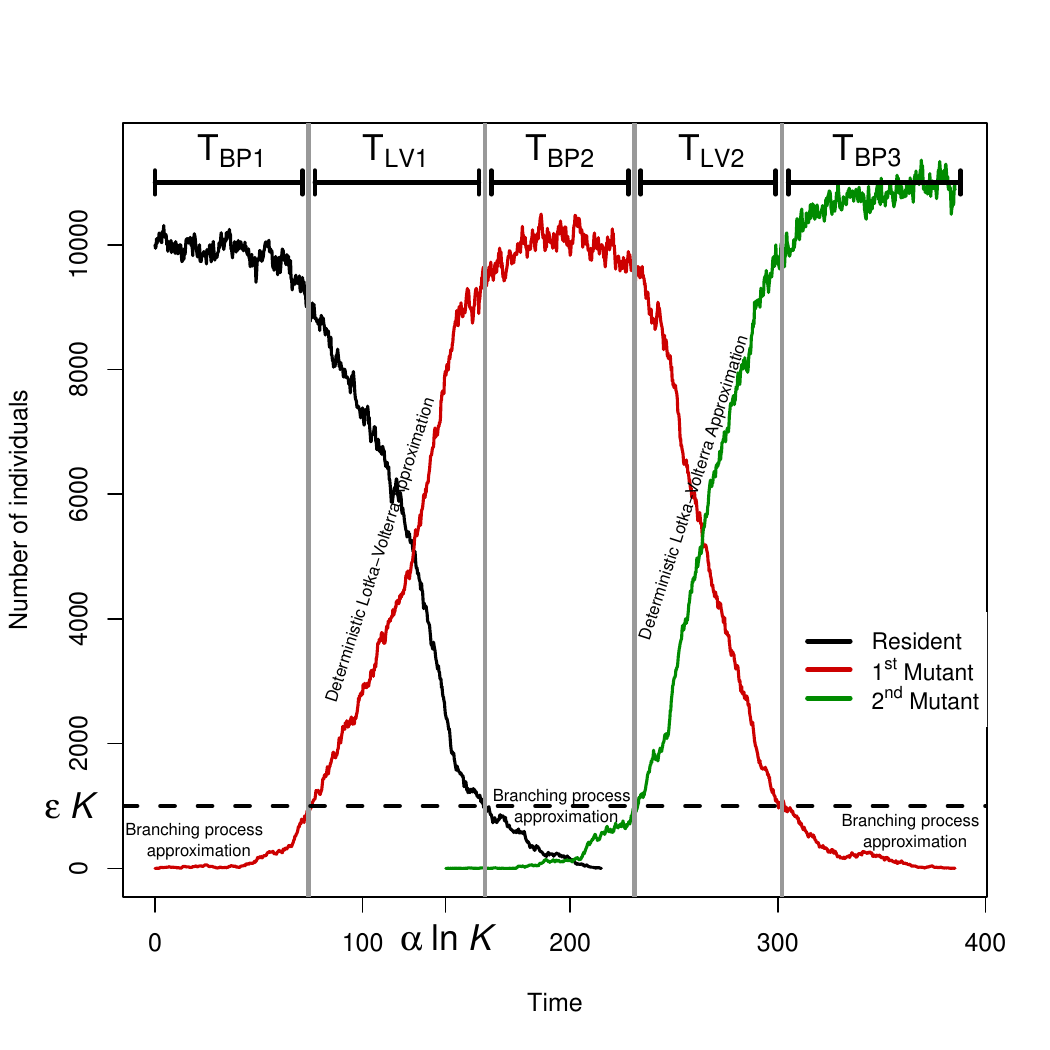}
%g b d h
\end{center}
\caption{Decomposition of the stochastic dynamics with three clones into a succession of five phases (delimited by the vertical grey lines). The black, 
red and green curves show the stochastic dynamics of the three clones in a simulation. Initially at time 0, the population is composed of the resident clone at 
stationary state (black curve)  and a single clone 1 individual is introduced into the population (red curve). After a time $\alpha \ln K$, the second mutant 
(green curve) is introduced into the population in a single copy. The resulting stochastic dynamics can be approximated by three supercritical branching processes 
(the first, third and fifth phases) and by two deterministic competitive Lotka-Volterra systems (the second and fourth phases). The phases approximated by 
the branching processes start when one or two clones have a population size lower than $\varepsilon K$ (dashed black line) while 
the other clones are at their stationary state. The phases approximated by the deterministic Lotka-Volterra system start 
when one of the clones has a population size larger than $\varepsilon K$. The duration of the five phases are shown at the top of figure : $T_{BP}$ and $T_{LV}$ respectively denote the duration of the phases approximated by the branching process and  by the Lotka-Volterra system. Estimates of the durations of each phase are given in the main text.}
\label{fig:phases}
%\end{sidewaysfigure}
\end{figure}

The dynamics of competing clones can be described by the succession of two kinds of phases, depending 
on the population size of each clone 
(Fig. \ref{fig:phases}):  either a phase approximated by a branching process without interactions (BP phase) or by a 
deterministic Lotka-Volterra model (LV phase). Hereafter, we describe more precisely these two approximations and  
what determines the succession of phases.

\par{\bf Approximation by a competitive Lotka-Volterra model.} If all clones populations are large, \ie of order $K$ with $K \rightarrow \infty$, the stochastic dynamics can be approximated by a three dimensions competitive Lotka-Volterra deterministic model (Fig. \ref{fig:phases}) \citep{fourniermeleard04, champagnat06, champagnatmeleard11,billiardsmadi17}. The variation of the density of each clonal strain is given by

\begin{equation} \label{EDO}
\left\{\begin{array}{ll}
	dn_0/dt=(\beta_0-\delta_0-C_{0,0}n_0-C_{0,1}n_1-C_{0,2}n_2)n_0\\
        dn_1/dt=(\beta_1-\delta_1-C_{1,0}n_0-C_{1,1}n_1-C_{1,2}n_2)n_1\\
        dn_2/dt=(\beta_2-\delta_2-C_{2,0}n_0-C_{2,1}n_1-C_{2,2}n_2)n_2
       \end{array} 
     \right.\\
      \text{ with } C_{ij} \geq 0 \text{ and } C_{ii}>0 \text{ for all } \{i,j\},
\end{equation}
where $n_i=N_i / K$ is the density of clone $i$. $N_i/K$ is the rescaled population size when $K$ is large, \ie the deterministic limit of the population size (note that we will keep the notation $N_i$ when dealing with stochastic dynamics and $n_i$ with deterministic dynamics). The model in Eq. \ref{EDO} can show different dynamics (stable fixed points or stable limit cycles)  and final states (monomorphic or polymorphic, with two or three coexistent clones), depending on the ecological parameters \cite[see][for details]{zeeman93,zeemanvandendriessche98,zeemanzeeman03}.

In order to provide approximations of the probabilities of invasion and fixation of a clone entering the population in a single copy, as well as its invasion and fixation times, the population size of each clone at the deterministic stable equilibrium is needed (see the definition of the \emph{invasion fitness} below). It is calculated thanks to Eq. \ref{EDO}. The population size of clone $j$ at equilibrium in a monomorphic population is given by $\bar{n}^j=\frac{\beta_j-\delta_j}{C_{jj}}$. $\bar{n}^j$ increases with the net reproductive rate $\beta_j-\delta_j$ and decreases with the intra-clonal competition intensity $C_{jj}$. If both clones $i$ and $j$ are present at equilibrium, the population sizes of clones $i$ and $j$ are given by 
\begin{equation} \label{eqpop}
\begin{array}{ll}\\
  \bar{n}^i_{ij}=\frac{C_{jj}(\beta_i-\delta_i)-C_{ij}(\beta_j-\delta_j)} {C_{ii}C_{jj}- C_{ij}C_{ji}}=\frac{\bar{n}^i - \frac{C_{ij}}{C_{ii}} \bar{n}^j}{1-\frac{C_{ij}C_{ji}}{C_{ii}C_{jj}}}, \qquad
  \bar{n}^j_{ij}=\frac{C_{ii}(\beta_j-\delta_j)-C_{ji}(\beta_i-\delta_i)}{C_{ii}C_{jj}- C_{ij}C_{ji}}=\frac{\bar{n}^j - \frac{C_{ji}}{C_{jj}} \bar{n}^i}{1-\frac{C_{ij}C_{ji}}{C_{ii}C_{jj}}}.
       \end{array}
     \end{equation}    
The rescaled size at equilbrium of clone $i$  in a dimporphic population  $\bar{n}^i_{ij}$ depends on the relative intensity between intra-clonal and inter-clonal competitions and the number of clone $j$ individuals $n_j$. The denominators in Eq. \ref{eqpop} show that the higher inter-clonal relative to intra-clonal competition, the higher the population size $\bar{n}^i$. 
     
\par{\bf Approximation of the stochastic dynamics by a branching process.} If at least one clone has a population size of order lower 
than $K$, the dynamics cannot be well approximated by a deterministic system, because it can be lost by chance 
(this loss is classically said to be due to demographic stochasticity or genetic drift). 
It is especially important when a clonal mutant $i$ enters in a resident $j$ population, or when clone $j$ is doomed to extinction after 
the invasion of clone $i$. The dynamics of a clone can however be approximated by a branching process as long as its population size is 
of order lower than $K$ (Fig. \ref{fig:phases}), and assuming competition between individuals $i$ is negligible relatively to competition 
from individuals $j$  \citep{fourniermeleard04, champagnat06, billiardsmadi17}.

The branching process approximating the dynamics of clone $i$ when rare has a growth rate $S_{ij}=\beta_i-\delta_i-C_{ij} \bar{n}^j$, which is generally called \emph{invasion fitness} of mutant $i$ in a resident population 
$j$ \citep{metzetal1996, fourniermeleard04}. If $S_{ij}>0$, clone $i$ is favored when rare in a resident $j$ population and can invade. 
Similarly, if the resident population is composed of two clones $i$ and $j$ at a steady state, then the fate of a clone $k$ entering in 
a single copy in the population is associated with the invasion fitness denoted 
$S_{kij}=\beta_k-\delta_k-C_{ki} \bar{n}^{i}_{ij}-C_{kj} \bar{n}^{j}_{ij}$. If $S_{kij}>0$, mutation $k$ is favorable when rare in the 
polymorphic resident population $(i,j)$ and can invade. The approximation by a branching process remains valid until the favored mutant 
$k$ reaches a population size of order $K$.

\par{\bf A succession of phases.} The sequence of succeeding phases only depends on the invasion fitness $S$, and on the time $\alpha \ln K$ when the second mutant enters the population.  Time is measured in $\ln K$ units because it is the relevant time scale under the intermediate regime defined before: the duration of invasion, fixation and extinction of a competing clone is of order $\ln K$ \citep{champagnat06, billiardsmadi17}. 
Assuming that the dynamics starts with the introduction of a single $i$ individual into a resident $j$ population, the dynamics thus starts with a BP phase (Fig. \ref{fig:phases}). The mutant $i$ invades the resident population $j$ \cite[with probability $S_{ij}/\beta_{i}$,][]{champagnat06, billiardsmadi17}. The population size of clone $i$ then becomes large, \ie of order $K$. 
A LV phase then starts, which can give different final states, only depending on invasion fitnesses 
$S_{ij}$ and $S_{ji}$ \citep{zeeman93,zeemanvandendriessche98,zeemanzeeman03}: either coexistence or one 
clone is doomed to extinction and reaches a size of order lower than $K$. In the latter case, a new BP 
phase starts which can end with either the loss of one clone, or with all clones having population sizes of order $K$. A new LV phase then starts, and so on. 

This succession of BP and LV  phases describes the dynamics of any number of competing clones \citep{billiardsmadi17}. Assuming we only consider three competing clones, two mutant clones $1$ and $2$ successively enter the resident population $0$ in a single copy. We focus only on cases where neither clones $0$ nor clones $1$ are lost when mutant $2$ appears, in particular we suppose that the invasive fitness of mutants $1$ and $2$ are positive when they appear. More 
precisely, we consider cases where $S_{10}>0$, and either i) $S_{20}>0$, if clone $2$ enters the population early when $1$ is still rare 
($\alpha \ln K$ is small enough, roughly $\alpha \ln K < 1/S_{10}$), or ii) $S_{21}>0$ or $S_{201}>0$ if mutation 2 enters the population 
when $1$ is common ($\alpha \ln K$ is large enough, roughly $\alpha \ln K > 1/S_{10}$).

 Since the ending state of phase $x$ is the initial state of phase $x+1$, only the initial condition of the first phase and the time $\alpha \ln K$ when mutant $2$ enters the population determine the sequence of succeeding phases, given that both 
clones $1$ and $2$ successfully invade the population.  When clone $2$ enters the population is important because either i) $\alpha \ln K$ is low enough 
\cite[roughly $\alpha \ln K < 1/S_{10}$,][]{billiardsmadi17} that clones $1$ are still rare and do not affect the invasion of mutants 
$2$, \ie only $S_{20}$ matters when $\alpha \ln K$ is lower than the time taken for clone $1$ to reach a large population size and the 
deterministic phase begins; Or ii) $\alpha \ln K$ is large enough that clones $1$ successfully invaded the population and thus affect 
the invasion of clones $2$, but clone $0$ is not lost \cite[roughly $ 1/S_{10} < \alpha \ln K < 1/S_{10}+1/{\mid S_{01} \mid} $,][]{billiardsmadi17}. In this case, only $S_{21}$ (if clones $0$ are rare) or $S_{201}$ (if clones $0$ are common) matters. 
In a nutshell, given the different invasion fitnesses $S$ and the time when mutant $2$ enters the population, it is possible to fully describe the different succeeding BP and LV phases until the final state is reached. The different possible final states are described in the next section.

\section*{Final states with three competing clones}\label{sec:final}

How the different final states and dynamics can be obtained are determined in six steps.
\begin{enumerate}

\item[{\bf Step 1.}] Does clone $2$ enter the population when $1$ is rare or common (depending on the time $\alpha \ln K$)? When clone 2 enters the population, it suffers from the competitive effect of clones $1$ or not, respectively when common or rare;

\item[{\bf Step 2.}] If clone $2$ enters the population when $1$ is rare, does clone $1$ or $2$ first reach the threshold population size  $\varepsilon K$? 
It depends on 
their invasion fitnesses $S_{10}$ and $S_{20}$ and on the time when mutant 2 enters the population $\alpha \ln K$. The first mutant which 
reaches a population size of order $K$ determines the initial state of the next LV phase;

\item[{\bf Step 3.}] What is the equilibrium of the first LV phase: stable coexistence of two clones or a single clone only? This only depends on the sign of the invasion fitnesses of both clones; 

\item[{\bf Step 4.}] What are the population sizes of all clones when the second BP phase begins? It depends on whether two clones stably coexist 
or not at the end of the previous LV phase (step 3.)?

\item[{\bf Step 5.}] Does a clone go extinct before the start of the next LV phase? When a clone has a population size of order lower than $K$ 
and is deleterious in a given context, it is expected to go extinct. However, its time to extinction can be longer than the time for 
another rare clone to reach the threshold population size $\varepsilon K$. In this case, a new LV phase begins. 

\item[{\bf Step 6.}] Steps 2-5 are  applied for the further successive phases (when applicable) as often as necessary until a final steady state is reached.
\end{enumerate}

Following this procedure \citep[see detailed computations in][]{billiardsmadi17}, Table \ref{tab:cases} summarizes the different possible 
final states for any competitive interactions, \ie all invasion fitness combinations, and for all times $\alpha \ln K$. Table \ref{tab:cases} shows that all final states are possible: fixation of $0$, $1$ or $2$, or the coexistence 
of all possible combinations between two or three clones. However, all final states do not occur under all conditions. When clone $2$ enters when $1$ is rare, only six final states are possible, whereas when it enters when $1$ is common seven final states can occur. 
Tab. \ref{tab:cases} also shows that the fixation of clone $0$, the fixation of  clone $1$ and the coexistence of $0$ and $1$ can only be obtained for a single set of conditions. All other final states can be obtained for various conditions. In particular, the fixation of clone $2$ can be obtained for very different conditions. As a consequence, the fixation time of clone $2$ is highly variable (see below).  

Assuming three competing clones, our model can thus capture a large diversity of dynamics, including one surprising final state: back to the initial state. Indeed, our model predicts that it is possible that clone $0$ goes to fixation even if the population is successively invaded by favorable clones $1$ and $2$. Interestingly, the Rock-Paper-Scissor dynamics is encountered under identical conditions regarding invasion fitness (Tab. \ref{tab:cases}). The Rock-Paper-Scissor and back to the initial state dynamics only differ by the time $\alpha \ln K$ when clone $2$ enters the population  (Tab. \ref{tab:cases}, Fig. \ref{fig:dynamics}c-d). This illustrates the importance of considering stochastic dynamics and the possible extinction of a given clone: if clone 2 enters the population late enough, clone 1 is lost before the invasion of clone 0, otherwise Rock-Paper-Scissors 
cyclical dynamics take place. In a deterministic model, for the same parameters, a mutant cannot go extinct and only Rock-Paper-Scissor dynamics 
are possible. Our results also show that Rock-Paper-Scissors dynamics can be obtained in a narrow set of conditions. Obviously, the first condition is that competitive interactions should be not transitive.  The second condition is less intuitive: the second clone should enter the population in a narrow time frame: it must occur after clone 1 invaded, since clone 2 is deleterious in a mutant 0 resident population. In addition, if clone 2 enters too late, then mutant 0 can be extinct before mutant 2 invades, in which case mutant 1 goes to fixation. These results have important consequences regarding our understanding of Rock-Paper-Scissor dynamics observed in natural populations: either the three types of individuals involved in such stable cycles have effectively entered the population by mutation or migration in a single copy, in which case the third type of individuals has necessarily entered the population in a narrow time frame. Otherwise, the alternative explanation is that the three types of individuals went together in a single population with a sufficiently large enough population size such that the dynamics initially followed an almost-deterministic dynamics, which certainly occurred by a massive migration and mixing of three different and complementary types of individual. \\ 

\section*{Invasion and fixation times} \label{sec:time}
Since a dynamics with two or three competing clones can be described by a succession of BP and LV phases, 
we can estimate the duration of invasion and fixation of a favorable mutant, \ie the time taken by the mutant clone to reach a 
population size of order $K$, and the time taken by the resident clones to go extinct, respectively. We give estimates of both invasion 
and fixation times in the case with two or three competing clones. Estimated times are compared with times obtained in stochastic 
individual based models (Simulations algorithm given in Appendix A2), for several population sizes (different values of $K$) and invasion fitnesses (different values 
of $S_{ij}$). 
 
\subsection*{Two clones}
When only two clones compete, the dynamics can be decomposed at most into three phases. The total duration 
of a dynamics with two competing clones can be decomposed in three times, corresponding to the three phases, 
denoted $T_{BP1}, T_{LV1}, T_{BP2}$, where the subscript $BP$ is used for a phase approximated by a branching process, and $LV$ 
for a phase approximated by a Lotka-Volterra deterministic system. 
The duration of the first BP phase is approximately, 
when $K \rightarrow \infty$, \cite[][Eq. 21, p.12]{durrett2015}
\begin{equation}\label{eq:Inv}
T_{BP1}=\frac{1}{S_{10}}\left(\ln\left(\varepsilon \ K \ \frac{S_{10}}{\beta_{1}} \right)+\gamma\right)
\end{equation}
where $\gamma \simeq 0.577$ is Euler's constant. Fig. \ref{fig:compar} shows the comparison between estimated and simulated invasion times for two competing clones with 
(arbitrarily chosen) $\varepsilon = 0.1$. Our results show that estimated and simulated invasion times are generally close, 
especially when $K$ is large enough.

The second phase follows a Lotka-Volterra dynamics for which there is no explicit formula of the time taken by a clone with size $\varepsilon$ to reach a size $(1-\varepsilon)\bar{n}_1$. However, 
we can roughly predict that this time is inversely proportional to $S_{10}$ and since the growth of clone 1 in a resident clone $0$ 
population is close to exponential at start, we can approximate the time by 
$T_{LV1}=1/S_{10} \ln((1-\varepsilon)/\varepsilon \  \bar{n}_1)$. 
Notice that $T_{BP1}+T_{LV1}$ is what is usually called 'fixation time' in population genetics \citep{desaifisher07}, as it 
corresponds to the time taken by the mutant to reach a large fraction of the population size. In our case, as the precise population 
composition is important to predict the fate of a new mutant, we also need to quantify the time taken by the clones $0$ to get extinct.
Assuming that at the end of the second phase, clone $0$ has a small population 
size, the competitive interactions suffered by clones $0$ are mostly due to clone $1$ 
($\varepsilon \ll C_{11}/C_{10} \bar{n}_1$) and the time to extinction of clone 0, \ie the duration of phase 3, is approximately given 
by $T_{BP2}= \ln K/|S_{01}|$.
The time of invasion of clone $1$ $T_{inv}$ is equal to the duration of the first phase $T_{BP1}$ while the time of fixation of 
clone $1$ is given by $T_{fix}=T_{BP1}+T_{LV1}+T_{BP2}$.  Figure \ref{fig:compar}a-b compares times of invasion $T_{inv}$ and fixation 
$T_{fix}$ obtained by stochastic simulations  and their approximations. Our results show that the approximations are generally
 close to simulations, at least of the same order. As expected, since the approximations are asymptotic when $K \rightarrow \infty$, 
the discrepancy is the largest for low population size (small $K$). Fig. \ref{fig:compar} also shows that the approximations are better 
for mutations with large effect (large $S_{ij}$). This is due to the fact that when $S_{ij}$ are low, stochasticity has a large effect 
and fixation times tend to be overestimated. Estimates are yet of the correct order of magnitude.

\subsection*{Three clones}
For the sake of simplicity, we estimate here the duration of the dynamics when clone $2$ eventually goes to fixation \cite[times of invasion and fixation for all cases can be obtained 
following similar calculations,][]{billiardsmadi17}. The dynamics can be decomposed into five successive phases with respective 
duration $T_{BP1}, T_{LV1}, T_{BP2}, T_{LV2}, T_{BP3}$, with a total duration of the dynamics given by $$ T_{fix}=T_{BP1}+T_{LV1}+ T_{BP2}+ T_{LV2}+T_{BP3}.$$ 
Clone $2$ enters the population either during the  first or the third phase of the dynamics (Fig. \ref{fig:phases}).

 When clone $2$ enters the population during the third phase ($T_{BP1}+T_{LV1}< \alpha \ln K$), the duration of the different phases can 
be estimated as if the dynamics were a succession of two independent dynamics with two competing clones only. Hence, the duration can be estimated using the same equations than in the previous 
section, with $T_{BP1}, T_{BP2}$ respectively the time of invasion of clone $1$ (resp. $2$) in a resident $0$ (resp. $1$) population, $T_{LV1}, T_{LV2}$ the duration of the two phases approximated 
by a Lotka-Volterra deterministic system, and $T_{BP3}$ the time taken for clone $1$ to get extinct. 

When clone $2$ enters the population during the first phase ($\alpha \ln K < T_{BP1}$), 
the duration of the third phase $T_{BP2}$ needs specific computations. Indeed, during the first phase, the clone 
$2$ population grows, and consequently the population size of clone $2$ at the beginning of the third phase must be taken into account. The duration of the third phase $T_{BP2}$ is then given by (see Appendix A1 for computation details)
\begin{equation}\label{eq:Inv2}
  T_{BP2}=\frac{1}{S_{21}} \left( 2 \gamma - \ln \left[ \beta_2 \exp(\gamma(\frac{S_{20}}{S_{10}}-1))K^{-S_{20} \alpha -1}\left( \frac{1}{\beta_1}K \bar{n}_1 S_{10} (1-\varepsilon )\right)^{S_{20}/S_{10}}\right] -\ln \left(S_{20} \varepsilon \right) \right).
\end{equation}
The durations of all other phases $T_{BP1}, T_{LV1}, T_{LV2}, T_{BP3}$ are not affected by the time at which clone $2$ enters 
the population. Figure \ref{fig:compar}c-d compares the estimated \emph{vs.} simulated fixation time of clone $2$ 
when it enters during the first phase and shows that our approximations are generally in good agreement when population size is large.

We compared the time taken for clone $2$ to invade a resident $0$ population with or without clone $1$. In other words, assuming $S_{10}>0$, $S_{20}>0$ and $S_{21}>0$, we can measure the interference effect of a favorable mutation $1$ on the fixation time of a favorable mutation $2$. For this, we compare $T^{*}_{inv}=\frac{1}{S_{20}}\left(\ln\left(\varepsilon \ K \ \frac{S_{20}}{\beta_{2}} \right)+\gamma\right)$ (Eq. \eqref{eq:Inv} in the case of clone $2$ invading a resident $0$ population without clone $1$) with $T_{inv}=T_{BP1}-\alpha \ \ln K + T_{LV1}+T_{BP2}$ (the duration of invasion of clone $2$ into a resident $0$ population when $1$ is present when clone $2$ enters at time $\alpha \ln K$). This comparison gives the following condition
\begin{equation}\label{eq:alpha}
  \scriptsize S_{10}(S_{21}-S_{20}) \alpha \ln K + \gamma(S_{20}-S_{10})< \ln \left[ \exp(\gamma(2+S_{21}/S_{10}-S_{21}/S_{20})) 
  \left(K \bar{n}_1 S_{10} (1-\varepsilon )/\beta_1\right)^{(S_{21}-S_{20})/S_{10}} \left( K S_{20} \varepsilon/\beta_2 \right)^{1-S_{21}/S_{20}}\right]  
\end{equation}
where clone $2$ invades faster a resident $0$ population when clone $1$ is present than when it is not. The inequality~\eqref{eq:alpha} means that, depending on invasion fitnesses  $S_{ij}$ and when clone $2$ enters the population, the interference between three clones can either slow down or speed up the invasion of a favorable mutation.

\section*{ Exploration of the parameter space assuming prior distributions}\label{sec:exploration}
%Dire ici que l'on va garder les estimations des temps pour K et S grands, et que cela ne devrait pas changer grand chose autre que les proportions relatives des différents cas. On peut discuter par exemple du fait que si K et S deviennent petits, sans doute que certains cas deviennent moins probables. De manières générale : on s'attend à ce que les dynamiques qui aient plus de succesion de phase soient moins reprentées parce que plus il y a de phase stochastique plus il y a de chance de perdre un des clones par hasard.

%Dire qu'on va explorer l'espace des paramètres (qui est grand voir tableau) pour voir si les régions de l'espace paramètrique dans lesquelles on observe telles ou telles dynamiques est grand ou pas (autrement dit est-ce vraisemblable ou pas ?).
We showed before that a large variety of dynamics can be observed when three clones compete, possibly with non-transitive competitive interactions. The dynamics can vary regarding their final states, or invasion and fixation times. Our model allows to predict each possible case given any set of ecological parameters and the time when clone $2$ enters the population (Tab. \ref{tab:cases} and Section \ref{sec:time}). How competition between three clones affects dynamics can be synthesized in six categories  (Tab. \ref{tab:cat}), depending on i) the final state: fixation or stable coexistence, ii) how invasion probability of clone $2$ is affected relatively to the case with only two interacting clones (compare the invasion fitnesses $S_{20}$ and $S_{21}$, for instance); iii) how fixation time of clone $2$ is affected (longer or shorter compared to the case with two competing clones, Eq. \ref{eq:alpha}). We introduce the following terms to describe the six possible categories of dynamics: ``clonal coexistence'', when clonal interaction promotes the maintenance of polymorphism; ``clonal assistance'' \emph{vs.} ``clonal hindrance'' when fixation time of clone $2$ is shortened or increased; ``soft'' \emph{vs.} ``hard'' when the invasion probability of clone $2$ is lower \emph{vs.} higher than with only two competing clones.

It is however difficult to have an overview of the likelihood of a particular dynamics, its final state and duration since the parameter space is very large, with many dimensions.  We now aim to explore the parameter space assuming prior distributions on the parameter space, and especially we aim at quantifying the 
likelihood of the different possible final states. The complexity of the model can be reduced by defining new parameters: $\rho_i=\beta_i-\delta_i$, the net individual reproductive rate of clone $i$, and $\widetilde{C}_{ij}=\frac{C_{ij}}{C_{jj}}$ the ratio of the between and within-clone competitive interactions. In bacteria, yeasts or some eukaryotes, fitness is generally estimated as the initial growth rate (at low density) of mutants (see Table 2 in \cite{martinlenormand06} and the Appendix in \cite{mannaetal12}). We thus assumed that the effect of mutations on the growth rate of clone $i$ follows a Fisher's geometric model. Given the net reproduction rate of clone 0 is $\rho_0$, we assumed that the reproductive rate of clone $i$ is $\rho_i=\rho_0 + x_i$ with $x_i$ the effect of mutation drawn in a shifted negative Gamma distribution (an approximation of a Fisher's geometric model \citep{martinlenormand06}).  Note that when mutation 2 enters the population during the second stochastic phase, mutation 2 is assumed to occur in the most frequent mutation at equilibrium: $\rho_2=\rho_1 + x_2$ when mutant 1 is more frequent than mutant 0, $\rho_2=\rho_0 + x_2$ otherwise.

There is, to our knowledge, no theoretical or empirical consensus on the distribution of mutation effects on the competitive abilities $\widetilde{C}_{ij}$. Without any knowledge about the distribution of competitive abilities, we simply assumed that the ratio of competitive interaction $\widetilde{C}_{ij}$ follows a uniform distribution in the interval $\left[1-u, 1+u \right]$, with $0 \leq u \leq 1$. Note that when $u=0$, all $\widetilde{C}_{ij}=1$, invasion fitnesses are necessarily transitive, while if $u>0$, non-transitivity can occur. As $u$ increases, the variance of the competitive ratio  $\widetilde{C}_{ij}$ also increases, i.e. the more different can the competitive interactions be between mutants. Finally, we assumed that the time at which clone $2$ enters the population $\alpha \ln K$ is uniformly distributed and occurs either during the first or third phase of the dynamics (Fig. \ref{fig:phases}). We randomly drew $10^6$ different sets of parameters in such prior distributions. For each parameter set, the final state was determined using Table \ref{tab:cases}. The posterior distribution of the final states was estimated as its proportion observed among the $10^6$ random parameters sets.

Figure \ref{fig:Distrib1} shows the posterior probability of the dynamics and final states when mutation 2 enters the population during the first or third phase of the dynamics, assuming that the competition abilities are drawn in a uniform distribution. When the variance of the distribution of the $\widetilde{C}_{ij}$ is small, all clones have similar competitive abilities ($\widetilde{C}_{ij} \simeq 1$), \ie invasion fitnesses are mostly transitive. We naturally recover predictions from population genetics models: The likeliest scenario is the fixation either of mutant 1 or 2 (Fig. \ref{fig:Distrib1}c, \ref{fig:Distrib1}d). Rapidly, when the variance of the uniform distribution increases, polymorphic final states become the likeliest. When the effect of mutation on competitive abilities becomes large ($u$ increases), the 
likelihood of all dynamics rapidly reaches a plateau. Our results suggest that non-transitive fitnesses are mostly expected to occur when several clones are interacting as soon as mutations affect their competitive abilities. This further supports that clonal coexistence is likely to occur even when considering only competitive interactions. Finally, our results show that Rock-Paper-Scissors dynamics and annihilation of adaptation are unlikely. Comparing left and right columns in Fig. \ref{fig:Distrib1} shows that the time at which clone $2$ enters the population only marginally affects the dynamics and the final states. Interestingly, comparing the final states between cases with two or three interacting clones (Fig.\ref{fig:Distrib2clones}) shows that more polymorphic final states are expected when three clones are interacting, even though the difference is small. Whether increasing the number of interacting clones could even more promote the maintenance of polymorphism is an open question.

Finally, Fig. \ref{fig:Distrib1}c-d shows the likelihood of clonal hindrance \emph{vs.} clonal assistance (\emph{sensu} Tab. \ref{tab:cat}).  Clonal hindrance is the most probable when the competitive abilities are similar between clones (small $u$). However, when the difference between competitive abilities increases (large $u$), 
the likelihood of clonal assistance increases. When clone $2$ enters the population during the third phase of the dynamics, clonal assistance is even likelier than clonal hindrance. Globally, our results thus suggest that clonal hindrance might indeed be an important factor affecting adaptation rate, but clonal assistance can be as important given non-transitive fitnesses are possible.

\section*{Discussion}

In this paper, we aimed at deciphering the dynamics of three competing clones. Despite its simplified assumptions, our model captures, 
at least qualitatively, all dynamics observed in evolution experiments: coexistence, fixation or extinction. Similar results were obtained in a model 
by \cite{goodetal2018} where a chemostat with several resources were assumed (\emph{i.e.} coexistence is possible because of niche differentiation), 
in a rare mutation limit (Regime A. in Fig.\ref{fig:regime}). However, this model does not allow non-linear and cyclical dynamics. 
In addition, unlike \cite{goodetal2018}, our model allows to estimate invasion and fixation times. It especially shows that the time when mutants enter into 
the population can dramatically affect the dynamics and the fate of the clones community. For instance, our model shows that Rock-Paper-Scissors dynamics can only 
take place by two successive mutation events only if the second mutation enters the population late enough. We also showed that competitive interactions between 
several clones can slow down or speed up invasion and fixation times, and can increase or decrease invasion probability of favorable mutations. Our results thus 
suggest that interference between several clones can affect adaptation in many different ways, and not necessarily only by slowing down adaptation rates, because 
of complicated ecological interactions that are potentially frequent in natural populations, such as intransitive competitive interactions. We introduced new 
terms describing how competitive interactions between several clones can affect adaptation (Table 3): clonal hindrance, assistance or coexistence when competitive 
interactions speed up or slow down fixation, or favor polymorphism. The effect of competitive interactions can be soft or hard, when it decreases or increases 
invasion probability. We argue that such a typology can help in better describing and understanding how clonal populations and communities evolve, especially in 
evolution experiments, by using concepts and vocabulary from both population ecology and population genetics.

 In the present work, contrarily to the literature dealing with clonal interference, we do not estimate adaptation rates. Indeed, we did not consider recurrent favorable mutations in the population and we did not suppose a particular distribution of mutational effects on fitness. Determining how adaptation rate is affected by clonal interference in a general context would necessitate further investigation. Stochastic dynamics of Lotka-Volterra models with more than three species could for instance be analyzed by numerical methods. Indeed, as demonstrated earlier by \cite{billiardsmadi17}, the stochastic dynamics can be well approximated by a succession of branching processes and deterministic ordinary differential equations. Hence, further investigating how interactions between several clones affect adaptation rates could be performed by combining i) general methods used to study the conditions for the invasion or the stability of coexistence of many species \cite[\eg][]{chesson2000, barabasetal2016, gallienietal2017}, ii) multi-type branching processes \cite[\eg][]{athreya2004}, and iii) supposing recurrent mutations (or immigrations) at random times with their effects on fitness drawn in particular distributions. In particular, it is well-known that the distribution of the effects of mutations on selection coefficients strongly affect adaptation rates \citep{neher13}. However, in theoretical studies, the effect of mutations are assumed to affect the selection coefficient, \ie implicitly the intrinsic growth rate, neglecting the effect of mutation on competitive interactions \citep{ desaifisher07, parketal2010, goodetal12}. How mutational effects on competitive interactions could affect adaptation rates is an open question.

Even though we did not study adaptation rates in the present paper, we explored how mutations on both the intrinsic growth rate \cite[supposed following a Fisher's adaptive landscape][]{martinlenormand06} and the competitive interactions between clones (\emph{a priori} supposed following a Uniform distribution) affect the final states of the population and invasion and fixation times. Our results showed that in agreement with deterministic models in the ecological literature, the ratio between intra and inter-species competition is the main factor affecting coexistence  \citep[\eg][]{chesson2000, chesson2018, barabasetal2016, gallienietal2017}. Our results show in particular that coexistence between two or three clones is very likely even for small mutational effects on competitive interactions (Fig. \ref{fig:Distrib1}, \ref{fig:Distrib2clones}). This suggests that the many cases of coexistence observed in evolution experiments can easily be explained by mutational effects on competitive interactions, even with very slight mutational effects. This prediction of our models calls for estimating competitive interactions between clones. Surprisingly, such experiments are scarce \citep{gallieni2017, friedmanetal2017} despite the existence of sophisticated experimental and statistical methods developed for this purpose \citep{taylorandaarssen1990, ulrichetal2014}. 

Our results highlight the importance of the time when the third clone enters the population. Table \ref{tab:cases} shows that this time can strongly affect the outcome of competition since in some cases, for fixed ecological parameters, coexistence is possible only when the appearance time is not too late or not too early. These results illustrate the importance of taking into account stochasticity when dealing with population and community dynamics especially when new species or strains enter the focal population or community with a 
low number of individuals. 
In other words, the final states of a community or population strongly isolated, \ie which receives rare immigrants or rare mutants, can behave very differently than non-isolated populations. Hence, investigating the effect of competitive interactions on coexistence only with deterministic models can have strong limitations. Interestingly, evolution experiments performed by \citet{hegrenessetal2006} indeed showed that the time of appearance of beneficial mutations can vary a lot between replicates. In their experiments, the relative abundance of two strains with different fluorescent markers is followed through time. The experiments outcomes vary a lot, corresponding to our model's predictions: either fixation or coexistence, or oscillating relative abundances. They showed that the most likely explanation for the variety of outcomes is not the difference in the mutational effects but rather the difference between the time of appearance of the beneficial mutations. These experiments' results support the predictions of our model that the time at which beneficial mutations occur into a population largely affect the outcome of competition between clonal strains.

We show that if mutations affect competitive interactions, interference between several clones can increase the probability of invasion of a favorable mutation, as well as speed up its invasion and fixation times (Fig. \ref{fig:Distrib1}, \ref{fig:Distrib2clones}). This suggests that considering clonal interference as slowing down adaptation rates only might not completely capture the effect of competition between several clones on adaptation rates, because intransitive competitive interactions can become important and make more complex ecological and evolutionary dynamics (also suggested by \cite{goodetal2018} in a chemostat with multiple resources model). Our model also predicts that, if intransitive competitive interactions are possible, then clonal assistance, \ie an acceleration of beneficial mutations fixation, is likely (Fig. \ref{fig:Distrib1}). Since long-term coexistence between several clonal strains has been observed in many different evolution experiments, clonal assistance is expected to frequently occur.

\cite{levyetal2015} estimated the mutational effects on fitness and the time of establishment (in generations) of mutations in a  large population of yeasts in a short-term experimental evolution ($\simeq 168$ generations). The estimates were obtained under the assumption that mutational effects on fitness were transitive. The observed dynamics of invading beneficial mutations show very similar patterns: a long phase of establishment followed by an exponential growth and finally a plateau. 
Surprisingly, despite the fact that a large number of favorable mutations coexist at the end of the experiments, the authors claimed that the observed dynamics is consistent without considering intransitive fitness interactions. In particular, they provide detailed predictions of the mutations that should be observed or not in their experiments, which are in perfect agreement with their observations (see Fig. 3a in \citet{levyetal2015}), which supports that the decrease in the fixation rates of beneficial mutations due to clonal interference is indeed a major mechanism underlying adaptation. The results in \citet{levyetal2015} challenge our own predictions that competitive interactions between several clones can be complex with either an increase or a decrease in the fixation of beneficial mutations. A possible explanation can be that indeed yeasts strains do not show intransitive competitive interactions, which could be tested experimentally. An alternative explanation can be that our model only considers three competing clones. \citet{grillietal2017} showed that increasing the number of competing species stabilizes their dynamics more rapidly because of higher order non-transitive interactions. Hence, the stable dynamics observed by \citet{levyetal2015} can be observed even considering intransitive interactions because of the coexistence of a large number of competing strains. To what extent the observed establishment times in \citet{levyetal2015} would also be in agreement with a stochastic model with more than three competing clones with intransitive competitive interactions is an open question.

\cite{goodetal2017} have different conclusions when analyzing clonal dynamics in the the long term evolution experiment with \emph{E. coli} (60 000 generations). 
By sequencing samples every 500 generations, they showed that the dynamics followed by each clonal strain was complex and very variable among the twelve replicates. Each replicate shows a succession of invasion and fixation of some clonal strains, as well as some phases with high polymorphism. Nine replicates among the twelve even show the coexistence of several strains during more than 10 000 generations, with sometimes what looks like cyclical dynamics. \cite{goodetal2017} estimated that the coexistence of several clonal strains could only be partly due to clonal interference. They concluded that other phenomena such as frequency-dependent selection (or intransitive competitive interactions) or ecological feedbacks should play an important role. Even if the dynamics in \cite{goodetal2017} are observed on a long-time scale, it suggests that the results of our model could explain the complexity of the dynamics of clonal populations, especially because we showed that long-term coexistence and clonal assistance are both very likely (their likelihood is of the same order than clonal hindrance, Fig. \ref{fig:Distrib1}). It is now needed to analyze our models into a larger time scale, close to the one in the long-term evolution experiment, in order to better disentangle the different mechanisms 
that underlie the adaptation of clonal populations.

As a conclusion, our present work illustrates that it is possible to bring together theoretical frameworks from population ecology and population genetics in order 
to have a better understanding of population, community and evolutionary dynamics. It is actually possible to integrate both ecological and genetic concepts into a 
single theoretical framework thanks to probabilistic mathematical tools such as the branching processes with interactions used here. We have shown that using a 
single model, it is possible to investigate at the same time the conditions for coexistence as well as probability and times of invasion and fixation of a 
beneficial mutation. Gathering several concepts into a single framework also highlights questions that are original in each field. For instance, on the one hand, 
investigating the conditions for coexistence for several competing species or strains by the use of deterministic models has many limitations and can give only a 
partial picture of the underlying mechanisms. Considering coexistence conditions under a stochastic framework is particularly important because most often new 
species enter communities as rare immigrants or mutants. On the other hand, estimating the invasion and fixation times of beneficial mutations assuming 
\emph{a priori} transitive fitness can also give a biased view of how adaptation is affected by the competition between several clones. This is particularly 
important because there are growing evidence that within-species non-transitive competitive interactions are widespread.

\section*{Acknowledgements}
\par We want to thank Guillaume Achaz and Guillaume Martin for helpful discussions. This work has been supported by the Chair ``Modélisation Mathématique et Biodiversité'' and was funded by the European Research Council (NOVEL project, grant n. 648321)

\bibliographystyle{amnat2}
\bibliography{reference}

\begin{thebibliography}{59}
\expandafter\ifx\csname natexlab\endcsname\relax\def\natexlab#1{#1}\fi
\expandafter\ifx\csname url\endcsname\relax
  \def\url#1{\texttt{#1}}\fi
\expandafter\ifx\csname urlprefix\endcsname\relax\def\urlprefix{URL }\fi

\bibitem[{Abelson and Loya(1999)}]{abelsonandloya1999}
Abelson, A. and Loya, Y.
\newblock 1999.
\newblock Interspecific aggression among stony corals in eilat red sea: a
  hierarchy of aggression ability and related parameters.
\newblock Bulletin of Marine Science 65:851--860.

\bibitem[{Ackerly and Cornwell(2007)}]{ackerlyandcornwell2007}
Ackerly, D. and Cornwell, W.
\newblock 2007.
\newblock A trait-based approach to community assembly: partitioning of species
  trait values into within- and among-community components.
\newblock EcolLett 10:135--145.

\bibitem[{Athreya and Ney(2004)}]{athreya2004}
Athreya, K.~B. and Ney, P.~E., 2004.
\newblock Branching processes.
\newblock Courier Corporation.

\bibitem[{Barab\`as et~al.(2016)Barab\`as, Michalska-Smith, and
  Allesina}]{barabasetal2016}
Barab\`as, G., Michalska-Smith, M., and Allesina, S.
\newblock 2016.
\newblock The effect of intra- and interspecific competition on coexistence in
  multispecies communities.
\newblock AmNat 188:E1--E12.

\bibitem[{Barrick and Lenski(2013)}]{barricklenski13}
Barrick, J. and Lenski, R.
\newblock 2013.
\newblock Genome dynamics during experimental evolution.
\newblock Nature Reviews Genetics 14:827--839.

\bibitem[{Behringer et~al.(2018)Behringer, Choi, Miller, Doak, Karty, Guo, and
  Lynch}]{behringeretal2018}
Behringer, M.~G., Choi, B., Miller, S.~F., Doak, T.~G., Karty, J.~A., Guo, W.,
  and Lynch, M.
\newblock 2018.
\newblock \emph{Escherichia coli} cultures maintain stable subpopulation
  structure during long-term evolution.
\newblock PNAS 115:E4642--E4650.

\bibitem[{Billiard and Smadi(2017)}]{billiardsmadi17}
Billiard, S. and Smadi, C.
\newblock 2017.
\newblock The interplay of two mutations in a population of varying size: a
  stochastic eco-evolutionary model for clonal interference.
\newblock Stochastic Processes and their Applications 127:701--748.

\bibitem[{Champagnat(2006)}]{champagnat06}
Champagnat, N.
\newblock 2006.
\newblock A microscopic interpretation for adaptive dynamics trait substitution
  sequence models.
\newblock Stochastic Processes and their Applications 116:1127--1160.

\bibitem[{Champagnat and M\'el\'eard(2011)}]{champagnatmeleard11}
Champagnat, N. and M\'el\'eard, S.
\newblock 2011.
\newblock Polymorphic evolution sequence and evolutionary branching.
\newblock Probability Theory and Related Fields 151:45--94.

\bibitem[{Chesson(2000)}]{chesson2000}
Chesson, P.
\newblock 2000.
\newblock Mechanisms of maintenance of species diversity.
\newblock ARES 31:343--366.

\bibitem[{Chesson(2018)}]{chesson2018}
Chesson, P.
\newblock 2018.
\newblock Updates on mechanisms of maintenance of species diversity.
\newblock Journal of Ecology 106:1773--1794.

\bibitem[{Crow and Kimura(1965)}]{crowkimura1965}
Crow, J.~F. and Kimura, M.
\newblock 1965.
\newblock Evolution in sexual and asexual populations.
\newblock The American Naturalist 99:439–450.

\bibitem[{de~Visser and Rozen(2006)}]{devisserrozen06}
de~Visser, J.~A.~G. and Rozen, D.~E.
\newblock 2006.
\newblock Clonal interference and the periodic selection of new beneficial
  mutations in \emph{Escherichia coli}.
\newblock Genetics 172:2093--2100.

\bibitem[{Desai and Fisher(2007)}]{desaifisher07}
Desai, M.~M. and Fisher, D.~S.
\newblock 2007.
\newblock Beneficial mutation-selection balance and the effect of linkage on
  positive selection.
\newblock Genetics 176:1759--1798.

\bibitem[{Doebeli(2002)}]{doebeli2002}
Doebeli, M.
\newblock 2002.
\newblock A model for the evolutionary dynamics of cross-feeding polymorphisms
  in microorganisms.
\newblock Population Ecology 44:59--70.

\bibitem[{Durrett(2015)}]{durrett2015}
Durrett, R., 2015.
\newblock Branching process models of cancer.
\newblock Springer International Publishing, Switzerland.

\bibitem[{Fisher(1930)}]{fisher1930}
Fisher, R.~A., 1930.
\newblock Plant breeding systems.
\newblock Oxford University Press, Oxford, UK.

\bibitem[{Fournier and Méléard(2004)}]{fourniermeleard04}
Fournier, N. and Méléard, S.
\newblock 2004.
\newblock A microscopic probabilistic description of a locally regulated
  population and macroscopic approximations.
\newblock The Annals of Applied Probability 14:1880--1919.

\bibitem[{Friedman et~al.(2017)Friedman, Higgins, and Gore}]{friedmanetal2017}
Friedman, J., Higgins, L.~M., and Gore, J.
\newblock 2017.
\newblock Community structure follows simple assembly rules in microbial
  microcosms.
\newblock Nature Ecology and Evolution 1:1--7.

\bibitem[{Gallieni(2017)}]{gallieni2017}
Gallieni, L.
\newblock 2017.
\newblock Intransitive competition and its effects on community functional
  diversity.
\newblock Oikos 126:615--623.

\bibitem[{Gallieni et~al.(2017)Gallieni, Zimmermann, Levine, and
  Adler}]{gallienietal2017}
Gallieni, L., Zimmermann, N., Levine, J., and Adler, P.
\newblock 2017.
\newblock The effects of intransitive competition on coexistence.
\newblock EcolLett 20:791--800.

\bibitem[{Gerrish and Lenski(1998)}]{gerrishlenski98}
Gerrish, P.~J. and Lenski, R.~E.
\newblock 1998.
\newblock The fate of competing beneficial mutations in an asexual population.
\newblock Genetica 102:127--144.

\bibitem[{Good et~al.(2018)Good, Martis, and Hallatschek}]{goodetal2018}
Good, B., Martis, S., and Hallatschek, O.
\newblock 2018.
\newblock Adaptation limits ecological diversification and promotes ecological
  tinkering during the competition for substitutable resources.
\newblock Proceedings of the National Academy of Sciences of the USA
  115:E10407--E10416.

\bibitem[{Good et~al.(2017)Good, McDonald, Barrick, Lenski, and
  Desai}]{goodetal2017}
Good, B., McDonald, M., Barrick, J., Lenski, R., and Desai, M.
\newblock 2017.
\newblock The dynamics of molecular evolution over 60,000 generations.
\newblock Nature 551:45--50.

\bibitem[{Good et~al.(2012)Good, Rouzined, Balick, Hallatschekg, and
  Desai}]{goodetal12}
Good, B.~H., Rouzined, I.~M., Balick, D.~J., Hallatschekg, O., and Desai, M.~M.
\newblock 2012.
\newblock Distribution of fixed beneficial mutations and the rate of adaptation
  in asexual populations.
\newblock Proceedings of the National Academy of Sciences of the U.S.A.
  109:4950--4955.

\bibitem[{Greaves and Maley(2012)}]{greavesmaley12}
Greaves, M. and Maley, C.~C.
\newblock 2012.
\newblock Clonal evolution in cancer.
\newblock Nature 481:306--313.

\bibitem[{Grilli et~al.(2017)Grilli, Barab\'as, Michalska-Smith, and
  Allesina}]{grillietal2017}
Grilli, J., Barab\'as, G., Michalska-Smith, M.~J., and Allesina, S.
\newblock 2017.
\newblock Higher-order interactions stabilize dynamics in competitive network
  models.
\newblock Nature 548:210--214.

\bibitem[{Hegreness et~al.(2006)Hegreness, Shoresh, Hartl, and
  Kishony}]{hegrenessetal2006}
Hegreness, M., Shoresh, N., Hartl, D., and Kishony, R.
\newblock 2006.
\newblock An equivalence principle for the incorporation of favorable mutations
  in asexual populations.
\newblock Science 311:1615--1617.

\bibitem[{Helling et~al.(1987)Helling, Vargas, and Adams}]{hellingetal87}
Helling, R.~B., Vargas, C.~N., and Adams, J.
\newblock 1987.
\newblock Evolution of \emph{Escherichia coli} during growth in constant
  environment.
\newblock Genetics 116:349--358.

\bibitem[{Kinnersley et~al.(2014)Kinnersley, Wenger, Kroll, Adams, Sherlock,
  and Rosenzweig}]{kinnersleyetal14}
Kinnersley, M., Wenger, J., Kroll, E., Adams, J., Sherlock, G., and Rosenzweig,
  F.
\newblock 2014.
\newblock \emph{Ex Uno Plures}: Clonal reinforcement drives evolution of a
  simple microbial community.
\newblock PLOS Genetics 10:e1004430.

\bibitem[{Lang et~al.(2011)Lang, Botstein, and Desai}]{langetal11}
Lang, G.~I., Botstein, D., and Desai, M.~M.
\newblock 2011.
\newblock Genetic variation and the fate of beneficial mutations in asexual
  populations.
\newblock Genetics 188:647--661.

\bibitem[{Lang et~al.(2013)Lang, D.~P.~Rice, Sodergren, Weinstock, Botstein,
  and Desai.}]{langetal13}
Lang, G.~I., D.~P.~Rice, M.~J.~H., Sodergren, E., Weinstock, G.~M., Botstein,
  D., and Desai., M.~M.
\newblock 2013.
\newblock Pervasive genetic hitchhiking and clonal interference in forty
  evolving yeast populations.
\newblock Nature 500:571--574.

\bibitem[{Levy et~al.(2015)Levy, Blundell, Petrov, Fisher, and
  Sherlock}]{levyetal2015}
Levy, S.~F., Blundell, J.~R., Petrov, S. V. D.~A., Fisher, D., and Sherlock, G.
\newblock 2015.
\newblock Quantitative evolutionary dynamics using high-resolution lineage
  tracking.
\newblock Nature 519:181--186.

\bibitem[{Llaurens et~al.(2017)Llaurens, Whibley, and Joron}]{llaurensetal2017}
Llaurens, V., Whibley, A., and Joron, M.
\newblock 2017.
\newblock Genetic architecture and balancing selection: the life and death of
  differentiated variants.
\newblock Molecular Ecology 26:2430--2448.

\bibitem[{Maddamsetti et~al.(2015)Maddamsetti, Lenski, and
  E.}]{maddamsettietal15}
Maddamsetti, R., Lenski, R.~E., and E., J.
\newblock 2015.
\newblock Adaptation, clonal interference, and frequency-dependent interactions
  in a long-term evolution experiment with \emph{Escherichia coli}.
\newblock Genetics 200:619--631.

\bibitem[{Maharjan et~al.(2006)Maharjan, Seeto, Notley-McRobb, and
  Ferenci}]{maharjanetal06}
Maharjan, R., Seeto, S., Notley-McRobb, L., and Ferenci, T.
\newblock 2006.
\newblock Clonal adaptive radiation in a constant environment.
\newblock Science 313:514--517.

\bibitem[{Manna et~al.(2012)Manna, Gallet, Martin, and Lenormand}]{mannaetal12}
Manna, F., Gallet, R., Martin, G., and Lenormand, T.
\newblock 2012.
\newblock The high-throughput yeast deletion fitness data and the theories of
  dominance.
\newblock Journal of Evolutionary Biology 25:892--903.

\bibitem[{Martin and Lenormand(2006)}]{martinlenormand06}
Martin, G. and Lenormand, T.
\newblock 2006.
\newblock A general multivariate extension of fisher’s geometrical model and
  the distribution of mutation fitness effects across species.
\newblock Evolution 60:893--907.

\bibitem[{May and Leonard(1975)}]{mayandleonard1975}
May, R. and Leonard, W.
\newblock 1975.
\newblock Nonlinear aspects of competition between three species.
\newblock SIAM Journal of Applied Mathematics 29:243--253.

\bibitem[{Metz et~al.(1996)Metz, Geritz, Meszéna, Jacobs, and
  Heerwaarden}]{metzetal1996}
Metz, J., Geritz, S., Meszéna, G., Jacobs, F., and Heerwaarden, J.~V.
\newblock 1996.
\newblock Adaptive dynamics, a geometrical study of the consequences of nearly
  faithful reproduction.
\newblock Stoch. Spat. Struct. Dyn. Syst. 45:183--231.

\bibitem[{Miralles et~al.(1999)Miralles, Gerrish, Moya, and
  Elena}]{mirallesetal99}
Miralles, R., Gerrish, P., Moya, A., and Elena, S.
\newblock 1999.
\newblock Clonal interference and the evolution of {RNA} viruses.
\newblock Science 285:1745--1747.

\bibitem[{Muller(1932)}]{muller32}
Muller, H.
\newblock 1932.
\newblock Some genetic aspects of sex.
\newblock The American Naturalist 8:118--138.

\bibitem[{Nahum et~al.(2011)Nahum, Harding, and Kerr}]{nahumetal2011}
Nahum, J.~R., Harding, B.~N., and Kerr, B.
\newblock 2011.
\newblock Evolution of restraint in a structured rock paper scissors community.
\newblock PNAS 108:10831--10838.

\bibitem[{Neher(2013)}]{neher13}
Neher, R.
\newblock 2013.
\newblock Genetic draft, selective interference, and population genetics of
  rapid adaptation.
\newblock Annual Review Ecology, Evolution and Systematics 44:195--215.

\bibitem[{Pacala and Tilman(1994)}]{pacalaandtilman1994}
Pacala, S. and Tilman, D.
\newblock 1994.
\newblock Limiting similarity in mechanistic and spatial models of plant
  competition in heterogeneous environments.
\newblock AmNat 143:222--257.

\bibitem[{Park et~al.(2010)Park, Damien, and Joachim}]{parketal2010}
Park, S.-C., Damien, S., and Joachim, K.
\newblock 2010.
\newblock The speed of evolution in large asexual populations.
\newblock Journal of Statistical Physics 138:381--410.

\bibitem[{Rainey and Travisano(1998)}]{raineyandtravisano1998}
Rainey, P.~B. and Travisano, M.
\newblock 1998.
\newblock Adaptive radiation in a heterogeneous environment.
\newblock Nature 394:69--72.

\bibitem[{Rosenzweig et~al.(1994)Rosenzweig, Sharp, Treves, and
  Adams}]{rosenzweigetal1994}
Rosenzweig, R.~F., Sharp, R.~R., Treves, D.~S., and Adams, J.
\newblock 1994.
\newblock Microbial evolution in a simple unstructured environment: Genetic
  differentiation in \emph{Escherichia coli}.
\newblock Genetics 137:903--917.

\bibitem[{Sinervo and Lively(1996)}]{sinervoandlively1996}
Sinervo, B. and Lively, C.
\newblock 1996.
\newblock The rock-paper-scissors game and the evolution of altermative male
  strategies.
\newblock Nature 380:240--243.

\bibitem[{Soliveres et~al.(2018)Soliveres, Lehmann, Boch, Altermatt, Carrara,
  Crowther, Delgado-Baquerizo, Kempel, Maynard, Rillig, Singh, Trivedi, and
  Allan}]{soliveresetal2018}
Soliveres, S., Lehmann, A., Boch, S., Altermatt, F., Carrara, F., Crowther,
  T.~W., Delgado-Baquerizo, M., Kempel, A., Maynard, D.~S., Rillig, M.~C.,
  Singh, B.~K., Trivedi, P., and Allan, E.
\newblock 2018.
\newblock Intransitive competition is common across five major taxonomic groups
  and is driven by productivity, competitive rank and functional traits.
\newblock Journal of Ecology 106:852--864.

\bibitem[{Taylor and Aarssen(1990)}]{taylorandaarssen1990}
Taylor, D. and Aarssen, L.
\newblock 1990.
\newblock Complex competitive relationships among genotypes of three perennial
  grasses: Implications for species coexistence.
\newblock AmNat 136:305--327.

\bibitem[{Tilman(1982)}]{tilman1982}
Tilman, D., 1982.
\newblock Resource Competition and Community Structure.
\newblock Princeton, NJ: Princeton Univ. Press.

\bibitem[{Ulrich et~al.(2014)Ulrich, Soliveres, Kryszewski, Maestre, and
  Gotelli}]{ulrichetal2014}
Ulrich, W., Soliveres, S., Kryszewski, W., Maestre, F.~T., and Gotelli, N.~J.
\newblock 2014.
\newblock Matrix models for quantifying competitive intransitivity from species
  abundance data.
\newblock Oikos 123:1057--1070.

\bibitem[{Vano et~al.(2006)Vano, Wildenberg, Anderson, Noel, and
  Sprott.}]{vanoetal06}
Vano, J.~A., Wildenberg, J.~C., Anderson, M.~B., Noel, J.~K., and Sprott.,
  J.~C.
\newblock 2006.
\newblock Chaos in low-dimensional {Lotka-Volterra} models of competition.
\newblock Nonlinearity 19:2391.

\bibitem[{Vetsigian(2017)}]{vetsigian2017}
Vetsigian, K.
\newblock 2017.
\newblock Diverse modes of eco-evolutionary dynamics in communities of
  antibiotic-producing microorganisms.
\newblock Nature Ecology and Evolution 1:1--9.

\bibitem[{Wang and Xiao(2010)}]{wangxiao10}
Wang, R. and Xiao, D.
\newblock 2010.
\newblock Bifurcations and chaotic dynamics in a 4-dimensional competitive
  lotka–volterra system.
\newblock Nonlinear Dynamics 59:411--422.

\bibitem[{Zeeman and Zeeman(2003)}]{zeemanzeeman03}
Zeeman, E. and Zeeman, M.
\newblock 2003.
\newblock From local to global behavior in competitive {Lotka-Volterra}
  systems.
\newblock Transactions of the American Mathematical Society 355:713--734.

\bibitem[{Zeeman(1993)}]{zeeman93}
Zeeman, M.~L.
\newblock 1993.
\newblock Hopf bifurcations in competitive three-dimensional lotka-volterra
  systems.
\newblock Dynamics and Stability of Systems 8:189--216.

\bibitem[{Zeeman and van~den Driessche(1998)}]{zeemanvandendriessche98}
Zeeman, M.~L. and van~den Driessche, P.
\newblock 1998.
\newblock Three-dimensional competitive lotka-volterra systems with no periodic
  orbits.
\newblock SIAM Journal on Applied Mathematics 58:227--234.

\end{thebibliography}
%\bibliography{biblio_clonal}

\newpage
\section*{Tables and Figures}

\begin{sidewaystable}[p!]
\begin{tabular}{c|ll}
  \textbf{\textit{Parameters}}   & \textbf{\textit{Definition}} & \textbf{\textit{Notes}}\\
  \textbf{\textit{and variables}} & & \\
  \hline
   $\beta_i$ & Reproduction rate of  a clone $i$ individual &  \\
 $\delta_i$ &  Intrinsic death rate of  a clone $i$ individual &  \\
  $C_{ij}$ & Competitive effect of a clone $j$ individual   &   Competition increases mortality,  hence $C_{ij} \geq 0$\\
  & $\qquad \qquad$ on  a clone $i$ individual  & \\
  $N_{i}$ &  Number of clone $i$ individuals &  \\
  $\bar{n}^i=\frac{\beta_i-\delta_i}{C_{ii}}$ & Population size at equilibrium with a single clone $i$ & Scaled with $K$\\
  $\bar{n}^i_{ij}=\frac{C_{ii}(\beta_j-\delta_j)-C_{ji}(\beta_i-\delta_i)}{C_{ii}C_{jj}-C_{ij}C_{ji}}$ & Population size at equilibrium of clone $i$ with clone $j$ & Scaled with $K$\\
  $S_{ij}=\beta_i-\delta_i-C_{ij} \bar{n}^j$ & Invasion fitness of clone $i$ into  &  If positive, clone $i$ has a probability $S_{ij}/\beta_i$ to invade, \\
               & $\qquad \qquad$ a resident $j$ population&  $\qquad \qquad$ zero otherwise (with $K\rightarrow \infty$)\\
  $S_{kij}=\beta_k-\delta_k-C_{ki} \bar{n}^i_{ij}-C_{kj} \bar{n}^j_{ij}$ & Invasion fitness of clone $k$ into  &  If positive, clone $k$ has a probability $S_{kij}/\beta_k$ to invade, \\
  & $\qquad \qquad$ a resident $i$ and $j$ population&  $\qquad \qquad$ zero otherwise (with $K\rightarrow \infty$)\\
  $K $ & Scaling parameter & Size of the system, especially gives the order  \\
  & &  $\qquad \qquad$ of the total population size\\
  $\alpha$ & Time when clone $2$ enters the population &  Time is measured in $ln K$ units\\
  $\varepsilon$ & Threshold size & Population size where dynamics can be approximated \\
               & & $\qquad \qquad$ by a deterministic Lotka-Volterra system\\
  $T_{BP}$, $T_{LV}$ & Duration of the phases of the dynamics approximated  & \\
               & $\qquad \qquad$ by a branching process or a Lotka-Volterra &  \\
  & $\qquad \qquad$ deterministic system&\\ 
\end{tabular}
\caption{Summary of parameters and variables.}\label{tab:summary}
\end{sidewaystable}

\newpage

\begin{sidewaystable}[p!]
\scriptsize
\begin{tabular}{c|ccc|ccc}
  & \multicolumn{6}{l}{  \textbf{\textit{Conditions for the different final states}}} \\
  \hline
  %& & & & & &\\
{ \textbf{\textit{Final}}}  & \multicolumn{2}{l}{ Clone 2 appears when clone 1 is rare (phase BP1)} & &\multicolumn{3}{l}{ Clone 2 appears when clone 1 is common (phase BP2)}  \\ 
%{ \textbf{\textit{Final}}} & & & & & &\\
%& \multicolumn{3}{l|}{(Clone 2 enters when 1 is rare)} & \multicolumn{3}{l}{(Clone 2 enters when 1 is common)}  \\ 
{ \textbf{\textit{States}}} & { $\alpha$}  & { $S>0$} & { $S<0$} & { $\alpha$} & { $S>0$} & { $S<0$}\\
  \hline
%& & & & & &\\
{ Fixation}         & $\emptyset$  & $\emptyset$  & $\emptyset$ & $ \frac{S_{02}}{|S_{12}| |S_{01}|}< \alpha -\frac{1}{S_{10}}+\frac{1}{S_{21}} <  \frac{1}{|S_{01}|} $&$S_{21}, S_{02}$ & $S_{01},S_{12},S_{20}$\\
{  0} & & & & &&\\
  \hline
  %& & & & & &\\
{ Fixation}          & & &$S_{01},S_{21}$ & & & $S_{01},S_{21}$\\
{ 1} &&&&&&\\
  \hline
  %& & & & & &\\
{ Fixation}         & &$S_{01}, S_{201}$ &$S_{12},S_{02}$ & &$S_{01}, S_{201},S_{20}$ &$S_{12},S_{02}$\\
{ 2}         &$\emptyset$ &     $\emptyset$           &      $\emptyset$        & &$S_{01}, S_{201},S_{21}$ &$S_{12},S_{02},S_{20}$\\
                  & $\alpha<\frac{1}{S_{10}}+\frac{1}{S_{20}}\left(\frac{S_{21}}{|S_{01}|}-1 \right)$&$S_{21}$ &$S_{01},S_{12},S_{02}$ &$\alpha -\frac{1}{S_{10}}+\frac{1}{S_{21}} <  \frac{1}{|S_{01}|}$ & &$S_{12},S_{02}$\\
                  &$\alpha>\frac{1}{S_{10}}+\frac{1}{S_{20}}\left(\frac{S_{21}}{|S_{01}|}-1 \right)$ & $S_{21}$ &$S_{01},S_{12}$ &$\alpha -\frac{1}{S_{10}}+\frac{1}{S_{21}} >  \frac{1}{|S_{01}|}$ & &$S_{12}$\\
& & & & & &\\
  \hline
  %& & & & & &\\
{ Coexistence}       & &$S_{01}$ &$S_{201}$ & &$S_{01}$ &$S_{201}$\\ 
{ 0 and 1} &&&&&&\\
  \hline
  %& & & & & &\\
{ Coexistence}        & &$S_{01},S_{201},S_{02},S_{21}$ &$S_{12},S_{102}$ & &$S_{01},S_{201},S_{20},S_{02},S_{21}$ & $S_{12}$\\
{ 0 and 2}         & &$S_{01},S_{201},S_{02},S_{12}$ &$S_{21},S_{102}$ & &$S_{01},S_{201},S_{20},S_{02},S_{12}$ & $S_{21}, S_{102}$\\
                   & &$S_{01},S_{201},S_{02}$ &$S_{21},S_{12}$ & &$S_{01},S_{201},S_{20},S_{02}$ & $S_{21}, S_{12}$\\
                   & &$S_{01},S_{201},S_{02},S_{12},S_{21}$ &$S_{102}$ & &$S_{01},S_{201},S_{20},S_{02},S_{12},S_{21}$ &$S_{102}$\\
                   & $\alpha<\frac{1}{S_{10}}+\frac{1}{S_{20}}\left(\frac{S_{21}}{|S_{01}|}-1 \right)$  &$S_{21},S_{12},S_{012}$ &$S_{01},S_{102}$ &$\alpha -\frac{1}{S_{10}}+\frac{1}{S_{21}} <  \frac{1}{|S_{01}|}$ & $S_{21},S_{12},S_{012},S_{20}$ & $S_{01},S_{102}$\\
                   & $\frac{S_{02}S_{21}}{|S_{12}||S_{01}|S_{10}}-1<\alpha<\frac{1}{S_{10}}+\frac{1}{S_{20}}\left(\frac{S_{21}}{|S_{01}|}-1 \right)$& $S_{21},S_{02}$&$S_{01},S_{12}$ & $ \frac{S_{02}}{|S_{12}| |S_{01}|}< \alpha -\frac{1}{S_{10}}+\frac{1}{S_{21}} <  \frac{1}{|S_{01}|} $&$S_{21},S_{02},S_{20}$ & $S_{01},S_{12}$ \\
  &$\alpha<\frac{S_{02}S_{21}}{|S_{12}||S_{01}|S_{10}}-1$ &$S_{21},S_{02}$ &$S_{01},S_{12},S_{102}$ & $ \alpha -\frac{1}{S_{10}}+\frac{1}{S_{21}} <\frac{S_{02}}{|S_{12}| |S_{01}|} $ & $S_{21},S_{02},S_{20}$ &$S_{01},S_{12},S_{102}$ \\
  & & & & & &\\
  \hline
  & & & & & &\\
{ Coexistence}    & & $S_{01},S_{201},S_{12},S_{21}$&$S_{012}$ & & $S_{01},S_{201},S_{12},S_{21}$ &$S_{02}$\\
{ 1 and 2}     &$\emptyset$ &$\emptyset$ &$\emptyset$ & & $S_{01},S_{201},S_{12},S_{21}$ & $S_{012}$\\  
                & $\alpha<\frac{1}{S_{10}}+\frac{1}{S_{20}}\left(\frac{S_{21}}{|S_{01}|}-1 \right)$& $S_{21},S_{12}$& $S_{01},S_{012}$ & $\alpha -\frac{1}{S_{10}}+\frac{1}{S_{21}} <  \frac{1}{|S_{01}|}$&$S_{21},S_{12}$ &$S_{01},S_{012}$\\
  & $\alpha>\frac{1}{S_{10}}+\frac{1}{S_{20}}\left(\frac{S_{21}}{|S_{01}|}-1 \right)$& $S_{21},S_{12}$ &$S_{01}$ &$\alpha -\frac{1}{S_{10}}+\frac{1}{S_{21}} >  \frac{1}{|S_{01}|}$ &$S_{21},S_{12}$ &$S_{01}$\\
  & & & & & &\\
  \hline
  %& & & & & &\\
{ Coexistence} & &$S_{01},S_{201},S_{12}$ &$S_{21},S_{02}$ & &$S_{01},S_{201},S_{20},S_{12}$ &$S_{21},S_{02}$\\
{ 0, 1 and 2}  & &$S_{01},S_{201},S_{12},S_{21},S_{012}$ & $S_{02}$ & &$S_{01},S_{201},S_{21},S_{02}$ &$S_{20},S_{12}$\\
            & &$S_{01},S_{201},S_{12},S_{02},S_{102}$ & $S_{21}$ & &$S_{01},S_{201},S_{20},S_{21},S_{12},S_{012}$ &$S_{02}$\\
            & &$S_{01},S_{201},S_{102},S_{21},S_{02}$ & $S_{12}$& &$S_{01},S_{201},S_{20},S_{21},S_{02},S_{102}$ &$S_{12}$\\
            & &$S_{01},S_{201},S_{12},S_{21},S_{02},S_{012},S_{102}$ & & &$S_{01},S_{201},S_{20},S_{02},S_{12},S_{102}$ &$S_{21}$\\
            &$\emptyset$ &$\emptyset$ &$\emptyset$ & &$S_{01},S_{201},S_{21},S_{02},S_{12},S_{012}$ & $S_{20}$\\
            &$\emptyset$ &$\emptyset$ &$\emptyset$ & &$S_{ij}, S_{ijk}, i,j,k\in \{0,1,2 \}$  &\\
            &$\alpha<\frac{1}{S_{10}}+\frac{1}{S_{20}}\left(\frac{S_{21}}{|S_{01}|}-1 \right)$ & $S_{21},S_{12},S_{012},S_{102}$ &$S_{01}$ & $\alpha -\frac{1}{S_{10}}+\frac{1}{S_{21}} <  \frac{1}{|S_{01}|}$ &$S_{21},S_{12},S_{012}$ & $S_{01},S_{20}$\\
            &$\emptyset$ &$\emptyset$ &$\emptyset$ &$\alpha -\frac{1}{S_{10}}+\frac{1}{S_{21}} <  \frac{1}{|S_{01}|}$ &$S_{21},S_{12},S_{012},S_{20},S_{102}$ &$S_{01}$\\
            & $\alpha<\frac{S_{02}S_{21}}{|S_{12}||S_{01}|S_{10}}-1$ &$S_{21},S_{02},S_{102}$  &$S_{01},S_{12}$ &$ \alpha -\frac{1}{S_{10}}+\frac{1}{S_{21}} <\frac{S_{02}}{|S_{12}| |S_{01}|} $ & $S_{21},S_{02},S_{20},S_{102}$ &$S_{01},S_{12}$\\
  { Rock-Paper}            &$\emptyset$ & $\emptyset$ &$\emptyset$ &$ \alpha -\frac{1}{S_{10}}+\frac{1}{S_{21}} <\frac{S_{02}}{|S_{12}| |S_{01}|} $ & $S_{21},S_{02}$ &$S_{01},S_{12},S_{20}$ \\
  -Scissor & & & & & &\\
\hline
\end{tabular}
\caption{Possible final states and their conditions. $\emptyset$: no condition; Blank space: for any parameter values}\label{tab:cases}
\end{sidewaystable}

\newpage
\begin{table}[p!]
\begin{tabular}{l|cccc}
&  &  {{\bf Lower}} & {\bf Higher}\\
 &   & {\bf invasion probability}  &  {\bf invasion probability}\\
 &   & ($S_{21}<S_{20}$ or $S_{201}<S_{20}$) &  ($S_{21}>S_{20}$ or $S_{201}>S_{20}$)\\
\hline
{\bf Polymorphism}  & &  Soft clonal coexistence & Hard clonal coexistence\\
  {\bf Fixation}  & & \\
  \hspace{1cm} - Faster fixation & &  Soft clonal assistance & Hard clonal assistance\\
  \hspace{1cm} - Slower fixation & &  Soft clonal hindrance & Hard clonal hindrance \\
\end{tabular}
\caption{Categories of possible dynamics.}\label{tab:cat}
\end{table}
\newpage
\newpage

%\begin{sidewaysfigure}[p!]
\newpage
%\begin{sidewaysfigure}
\begin{figure}
\begin{center}
\subfloat{
\includegraphics[trim = 0mm 5mm 10mm 20mm, clip, scale=0.55]{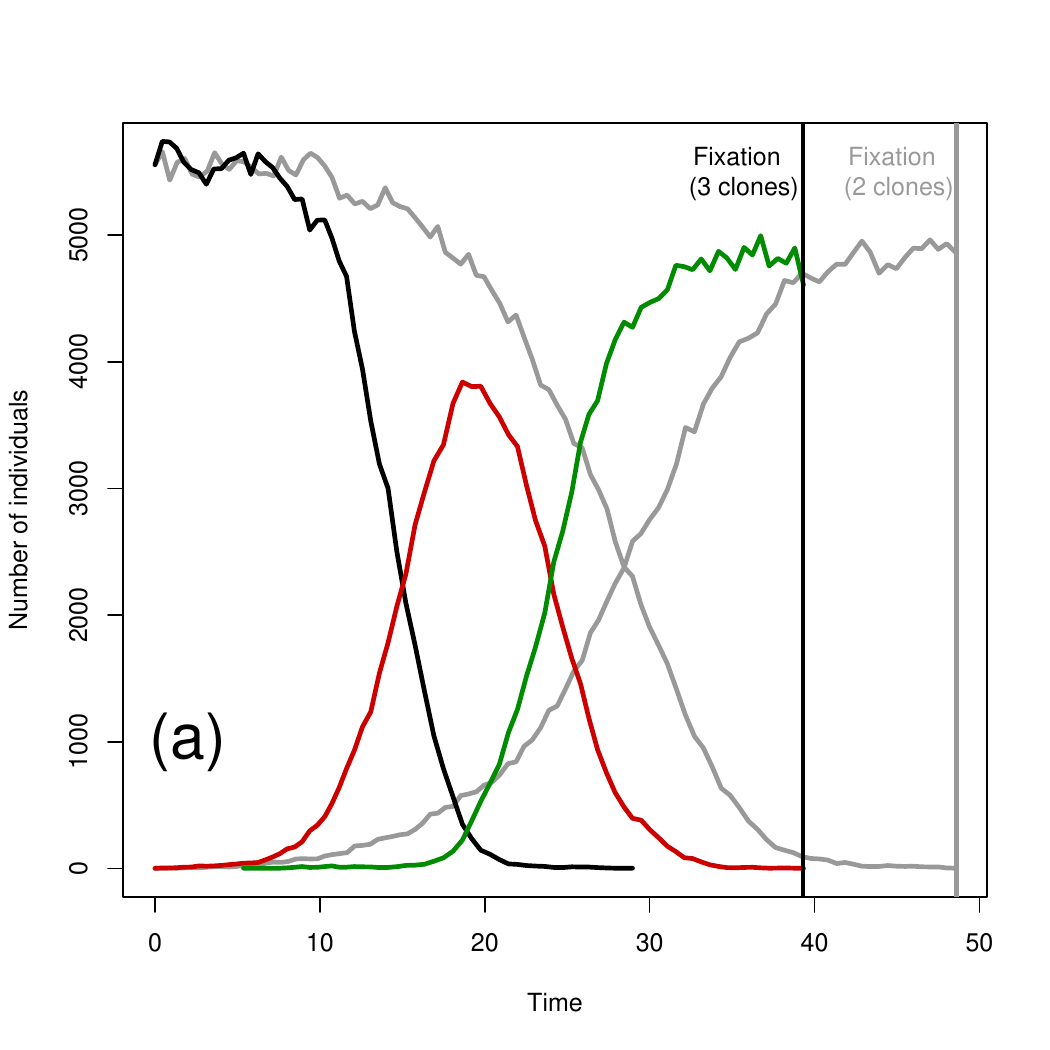}}
\subfloat{
\includegraphics[trim = 0mm 5mm 10mm 20mm, clip, scale=0.55]{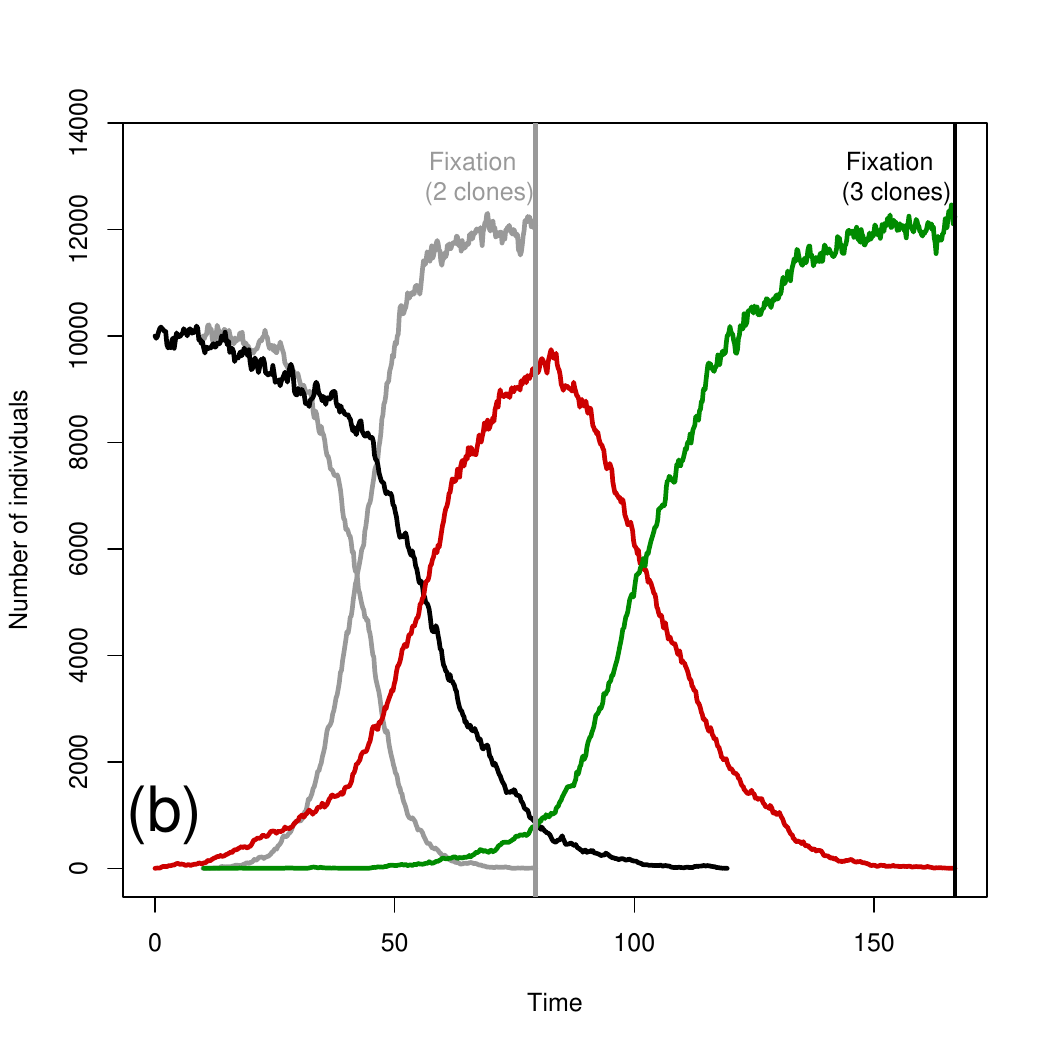}}\\
\subfloat{
\includegraphics[trim = 0mm 5mm 10mm 20mm, clip, scale=0.55]{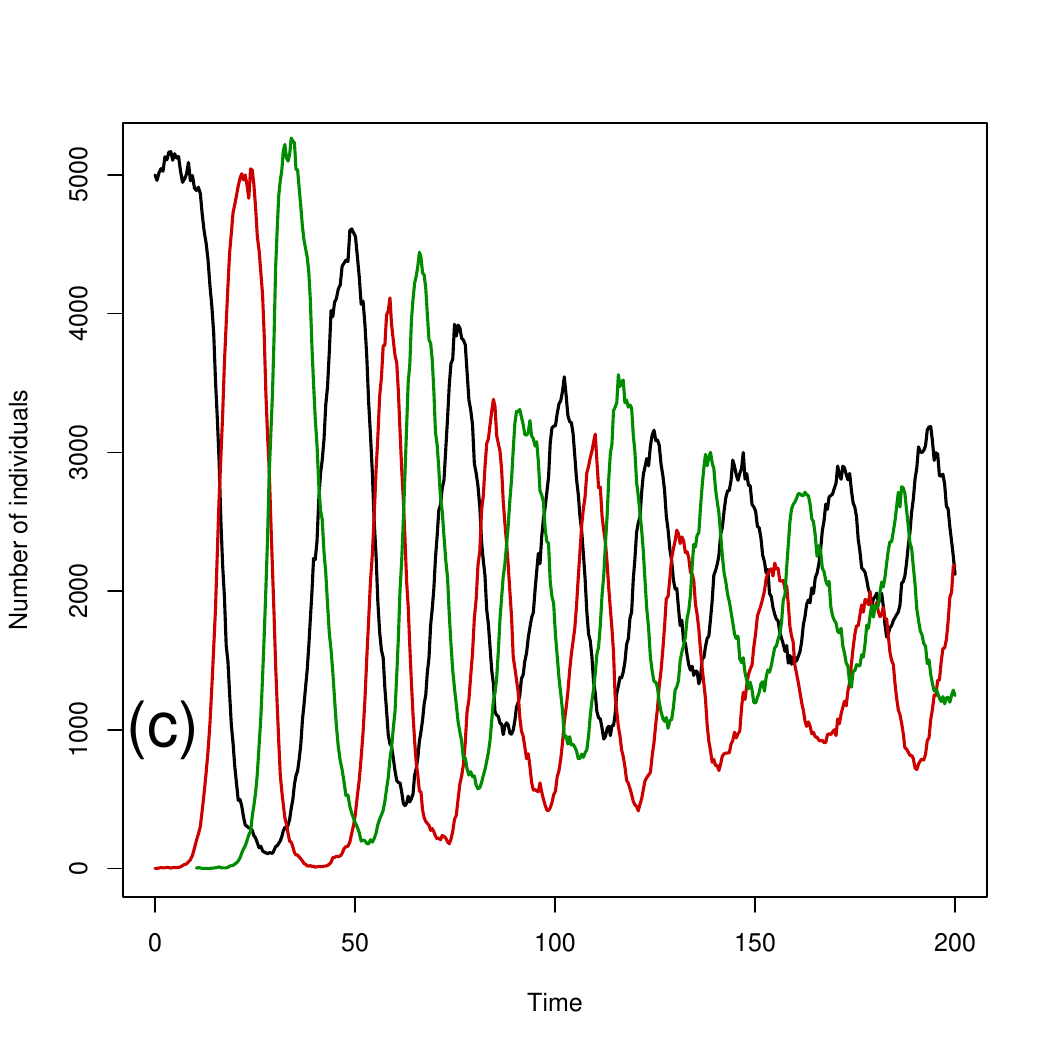}}
\subfloat{
\includegraphics[trim = 0mm 5mm 10mm 20mm, clip, scale=0.55]{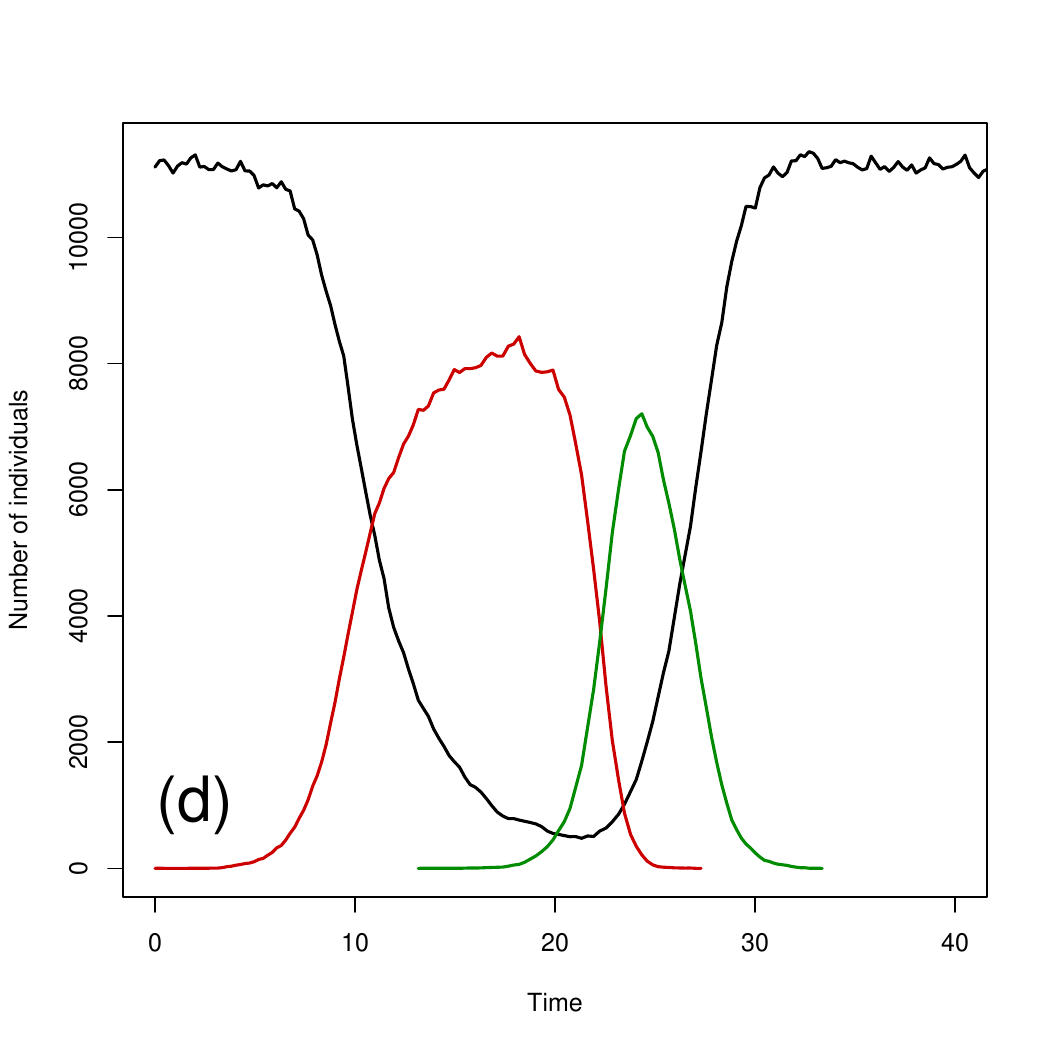}}
\end{center}
\caption{Different possible dynamics: simulations of the stochastic process. Black: Resident clone; red: clone $1$, green: clone $2$. (a) Clonal assistance: 
fixation of clone $2$ is faster with clone $1$ than without (compare gray and colored curves); (b) Clonal hindrance: fixation of clone $2$ is slower with 
clone $1$ than without (compare gray and colored curves); (c) An example of clonal coexistence: Rock-Paper-Scissor dynamics; (d) Annihilation of adaptation: 
Clone $0$ gets finally fixed after the successive invasions of clones $1$ and $2$. Parameters: $K=10000$; (a): $\beta_i=2$ and $\delta_i=1$ for all $i$, $C_{00}=1.8$,$C_{10}=C_{21}=1$, $C_{11}=2.3$,$C_{12}=3$, $C_{20}=1.5$, $C_{01}=4$, $C_{02}=3$, $C_{22}=2.1$, $\alpha \ln K = 5$; (b):  $C_{ij}=1$, $\alpha \ln K= 1.1$, $\beta_0=2$, $\beta_1=2.1$, $\beta_2=2.2$ and $\delta_i=1$ for all $i$, $\alpha \ln K = 10$; (c) $\beta_i=2$ and $\delta_i=1$ for all $i$, $\alpha \ln K = 10.1$, $C_{00}=C_{11}=C_{22}=2$, $C_{01}=2.5$, $C_{02}=C_{10}=C_{21}=1$, $C_{12}=C_{20}=3$; (d) $\beta_i=2$ and $\delta_i=0$ for all $i$, $\alpha \ln K = 13$, $C_{00}=1.8$, $C_{01}=2.5$, $C_{02}=1.5$, $C_{10}=C_{21}=1.0$, $C_{11}=2.3$, $C_{12}=5$, $C_{20}=3$, $C_{22}=2.1$.  }
\label{fig:dynamics}
\end{figure}
%\end{sidewaysfigure}

\newpage
\begin{figure}
\begin{center}
  \subfloat{\includegraphics[trim = 0mm 0mm 1mm 0mm, clip,scale=0.6]{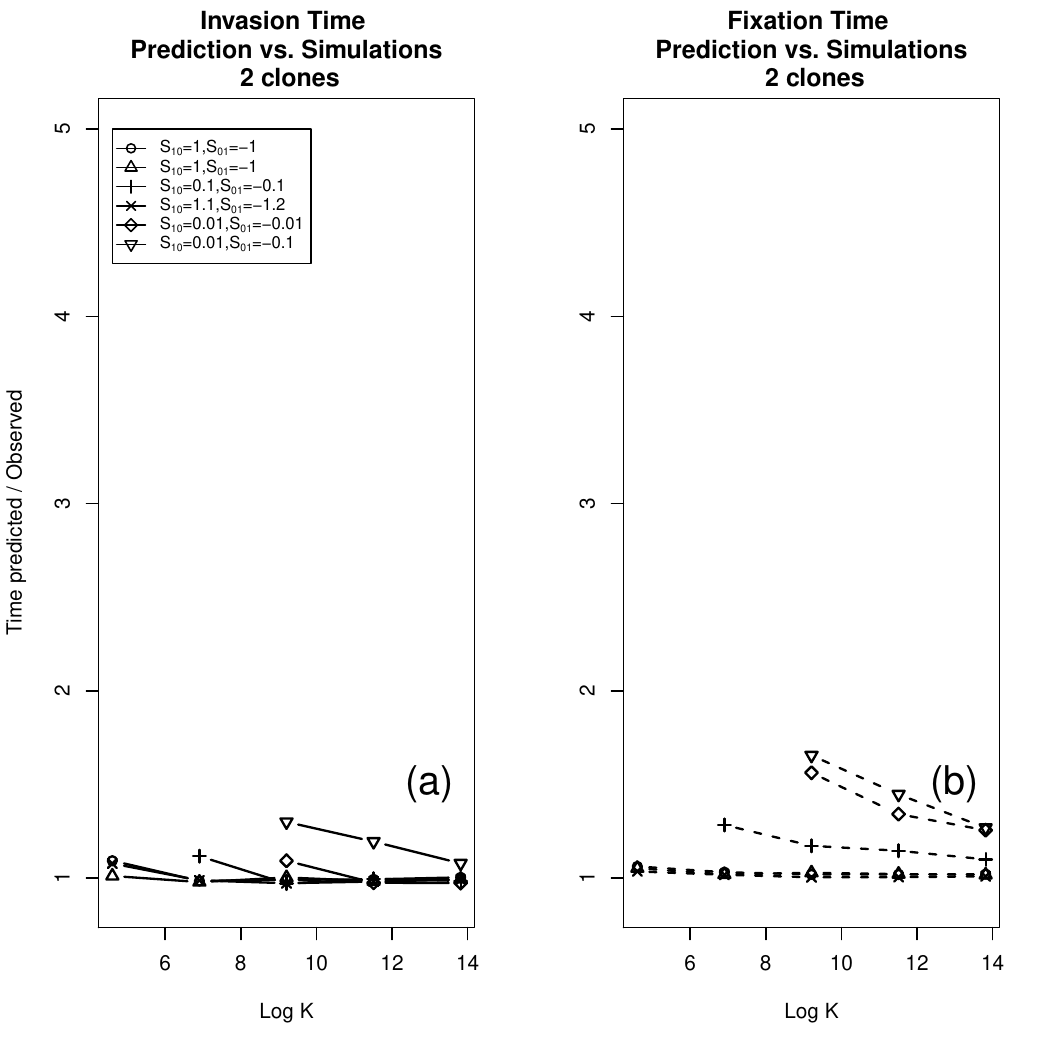}}\\

 \subfloat{\includegraphics[trim = 0mm 0mm 1mm 0mm, clip,scale=0.6]{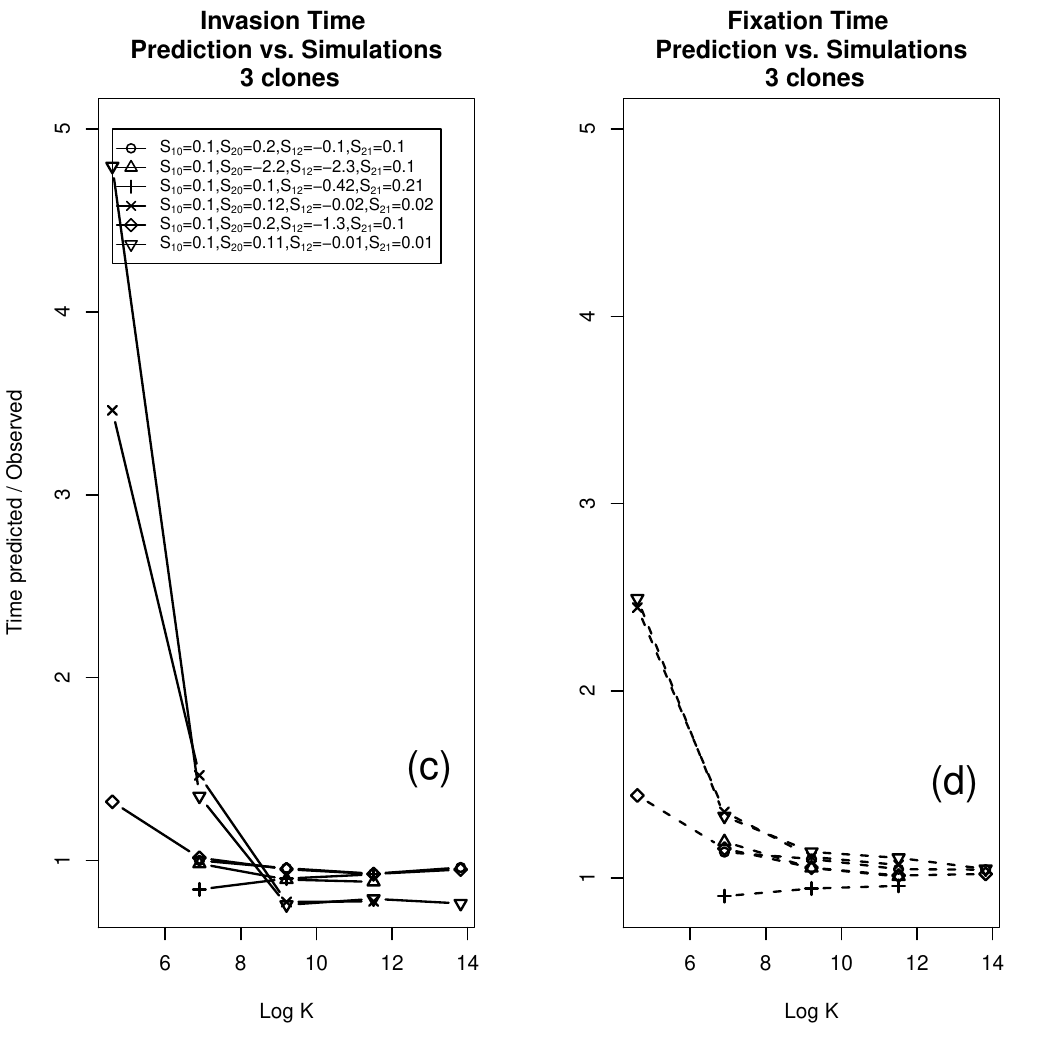}}
 \end{center}
\caption{Ration of estimated \emph{vs.} simulated invasion and fixation times with two or three competing clones. Top and bottom figures: two and three clones, respectively. Left and right figures: invasion or fixation times, respectively. The different symbols show different parameter sets (see legend). Estimated \emph{vs.} simulated times must be compared with 1: Above, the model's predictions (see main text) overestimate invasion or fixation times.  }
\label{fig:compar}
\end{figure}

\newpage
\begin{figure}
\begin{center}
\includegraphics[trim = 2mm 0mm 1mm 0mm, clip,scale=0.5]{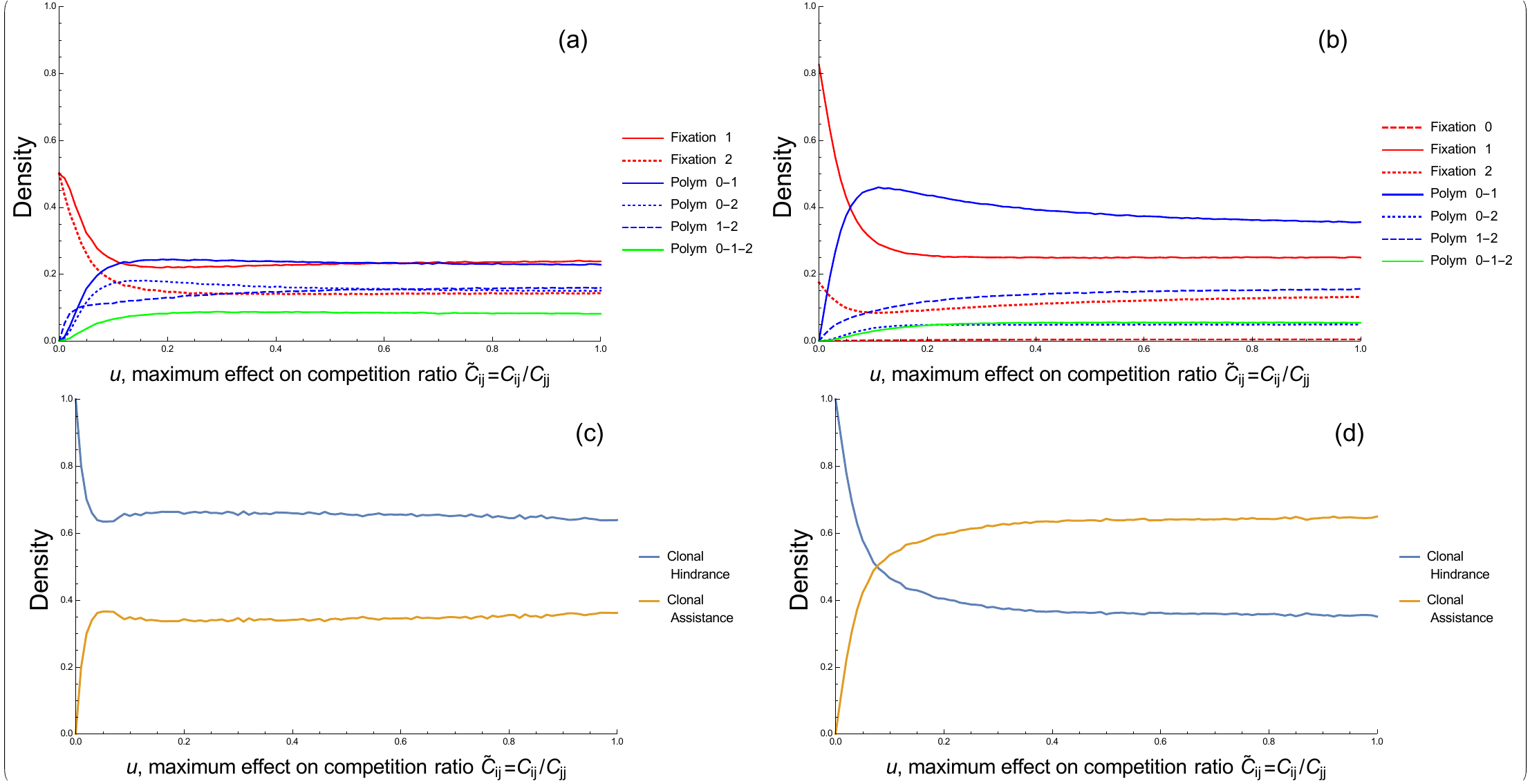}
\end{center}
\caption{ Posterior probability when competition coefficient is drawn in a uniform distribution. Left and right columns: clone $2$  enters in the first or third 
phase of the dynamics (Fig. \ref{fig:phases}), respectively.  (a)-(b) Final states;  (c)-(d) Clonal hindrance \emph{vs.} clonal assistance. The growth rate 
of clone $0$ $\rho_0=2$ is supposed to be at 50\% from the optimum in a Fisher's Geometric adaptive landscape (see text for details). The ratio of competitive 
abilities $\widetilde{C}_{ij}=C_{ij}/C_{jj}$ between clones $i$ and $j$ are drawn in a uniform distribution with range $\left[1-u, 1+u \right]$.}
\label{fig:Distrib1}
\end{figure}

\newpage
\begin{figure}
\begin{center}
\includegraphics[trim = 2mm 0mm 1cm 0mm, clip,scale=0.5]{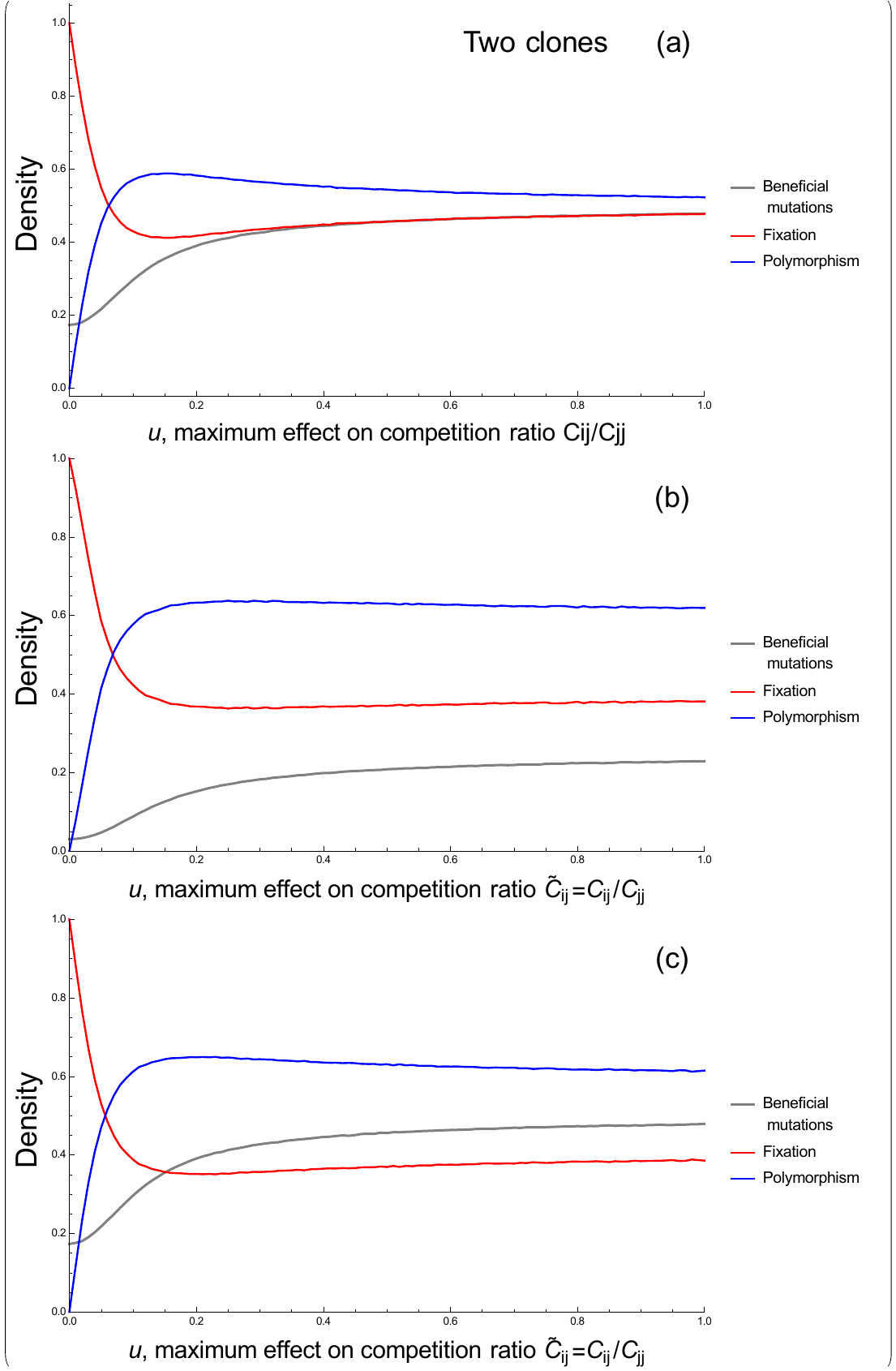}
\end{center}
\caption{ Polymorphism \emph{vs.} fixation when (a) there  are only two interacting clones (clones $0$ and $1$), or (b)-(c) three interacting clones (clones $0$, $1$ 
and $2$). The proportion of beneficial mutations among all randomly drawn parameters sets is also shown; Clone 2 enters the population either during the first phase 
(b) or third phase (c) of the dynamics.}
\label{fig:Distrib2clones}
\end{figure}

\newpage
\clearpage
\section*{Appendix}
\renewcommand{\theequation}{A\arabic{equation}}
% redefine the command that creates the equation number.
\renewcommand{\thetable}{A\arabic{table}}
\setcounter{equation}{0}  % reset counter 
\setcounter{figure}{0}
\setcounter{table}{0}

\subsection*{Appendix A1. Invasion and fixation times with three clones}\label{app:Tinv2}
In this appendix, our goal is to give an approximation for the invasion and fixation times of clone 2 in cases illustrated in Fig \ref{fig:dynamics}a-b. More precisely clone 2 enters the population during the first phase of the dynamics (\ie under the assumption that $\alpha \ln K < T_{BP1}$, see Eq. \eqref{eq:Inv}),  clone 1 reaches first the threshold $\varepsilon \ K$ (\ie such that $T_{BP1}<\frac{1}{S_{20}}\left(\ln\left(\varepsilon \ K \ \frac{S_{20}}{\beta_{2}} \right)+\gamma\right) +\alpha \ln K$), clone 2 invades the population after clone 1 and eventually goes to fixation. The duration of the dynamics with three clones can be splitted into five phases denoted $T_{BP1}, T_{LV1}, T_{BP2}, T_{LV2}, T_{BP3}$. $T_{BP1}$ and $T_{LV1}$ are the same than with only two clones. $T_{LV2}$ and $T_{BP3}$ can be calculated analogously: $T_{LV2}=1/S_{21} \ln((1-\varepsilon)/\varepsilon \  \bar{n_2})$ and $T_{BP3}=\frac{1}{|S_{12}|} \ln K$. The duration $T_{BP2}$ needs specific computations since the population size of clone 2 at the beginning of this phase depends on the duration $T_{BP1}$ and $T_{LV1}$: the larger the initial population size of clone 2, the shorter the time to reach the threshold size $\varepsilon K$. 

Let us denote $x$ the population size of clone 2 at the beginning of phase $BP2$. Assuming that the resident clone 0 is much more abundant than clones 1 and 2, competition on clones 2 is mostly due to clone 0, and we have $T(x)=\ln(x K S_{20}/\beta_{2}) +\gamma /S_{20}$ from \cite[][Eq. 21, p.12]{durrett2015}. Making this assumption neglects the increasing competitive interaction of clone 1 on clone 2, as well as the decreasing competitive interaction of clone 0 on clone 2 during phase $LV1$. This approximation should be correct when competitive effects of clones 0 and 1 on clone 2 are not too different (\ie $C_{20}\simeq C_{21}$ and $\bar{n}_0 \simeq \bar{n}_1$). Since clone 2 enters the population at time $\alpha \ln K$, $x$ is calculated by solving $T(x)=T_{BP1}+T_{LV1}-\alpha \ln K$, which gives
\begin{equation}\label{eq:x}
  x=\frac{1}{S_{20}} \frac{1}{K^{1+\alpha S_{20}}}\beta_2 \ \exp(\gamma (S_{20}/S_{10} -1)) \left( \frac{1}{\beta_1} (1-\varepsilon) K S_{10} \bar{n}_1 \right)^{S_{20}/S_{21}}.
\end{equation}
$T_{BP2}$ is finally given by the difference between the durations for a single clone 2 individual i) to reach the threshold $\varepsilon K$ when clone 1 is resident ($\ln(\varepsilon K S_{21}/\beta_{2}) +\gamma /S_{21}$) and ii) to reach a population size $x \ K$ ($\ln(x K S_{21}/\beta_{2}) +\gamma /S_{21}$). Replacing $x$ by Eq. \eqref{eq:x} gives
\begin{equation}
  T_{BP2}=\frac{1}{S_{21}} \left( 2 \gamma - \ln \left[ \beta_2 \exp(\gamma(\frac{S_{20}}{S_{10}}-1))K^{-S_{20} \alpha -1}\left( \frac{1}{\beta_1}K \bar{n}_1 S_{10} (1-\varepsilon )\right)^{S_{20}/S_{10}}\right] -\ln \left(S_{20} \varepsilon \right) \right).
\end{equation}
The estimation of the invasion and fixation times of clone 2, respectively given by $T_{BP1}+ T_{LV1}+ T_{BP2}$ and  $T_{BP1}+ T_{LV1}+ T_{BP2}+T_{LV2}+ T_{BP3}$ are compared with exact individual-based simulations in Fig. \ref{fig:compar}. (see simulation algorithm in App. A2).

\subsection*{Appendix A2. Simulation algorithm}\label{app:simu}
At time $T$, the total rate of possible events is given by $$\psi(T)=\sum_{i=0}^2 \left( \beta_i+\delta_i N_i(T)+\sum_{j}C_{ij}N_j(T) \right) N_i(T)$$ where $N_i(T)$ is the number of clone $i$ individuals in the population at time $T$, $b_i$ is the individual reproduction rate of clone $i$ individuals, $\delta_i$ the individual death rate, and $C_{ij}$ is the effect of competition of a single clone $j$ individual on a clone $i$ individual, affecting death.  The probability that at time $T+\Delta T $, the next event is a birth (resp. a death) of a clone $i$ individual is given by $\beta_i N_i(T)/ \psi(T)$ (resp. $d_i(N(T)) N_i(T)/\psi(T)$ where $d_i(N_i)=\delta_i N_i + \sum_j C_{ij}N_j(T))$. The time $\Delta T$ is drawn in an exponential distribution with parameter $\psi(T)$. If an individual $i$ is born (resp. is dead) then the size of the population of mutants $i$ becomes $N_i(T+\Delta T) = N_i(T)+1$ (resp. $N_i(T+\Delta T) = N_i(T)-1)$. The succession of events and the time taken for each event are randomly drawn until the desired final state is reached. Simulations were run either for illustrative purpose and show the different possible dynamics or to estimate invasion and fixation times. In the latter case, 200 independent replicates were run and the mean time among the replicates were calculated as an estimate of times. Note that we did not use this stochastic algorithm to explore the parametric space using prior distributions of the parameters. We determined the different final states  using Tab. \ref{tab:cases} given the ecological parameters summarized in $S_{ij}$, and $\alpha$ the time of appearance of mutation 2.

\clearpage
%\bibliographystyle{model2-names}
%\bibliography{reference.bib}
\end{document}